\documentclass[12pt,draftclsnofoot,onecolumn]{IEEEtran}
\makeatletter
\newcommand\semihuge{\@setfontsize\semihuge{22.3}{22}}
\makeatother
\usepackage{algpseudocode}
\usepackage{algorithm}
\usepackage{algorithmicx}
\usepackage{setspace}

\usepackage{lipsum} % For algorithm smape after or before:
\usepackage{amsfonts}

\usepackage[algo2e]{algorithm2e}
\usepackage[dvips]{color}
\usepackage{comment}
\usepackage{todonotes}
\usepackage{epsf}
\usepackage{epsfig}
\usepackage{bigints}
\usepackage{times}
\usepackage{epsfig}
\usepackage{graphicx}
\usepackage{mathtools}
\usepackage{mathrsfs}
\usepackage{amssymb}
\usepackage{pdfpages}
\usepackage{epstopdf}
\usepackage{float}
\newfloat{algorithm}{t}{lop}
\usepackage{subfig}
\usepackage[thinc]{esdiff}
\usepackage{commath}
\usepackage{mathtools}
\DeclarePairedDelimiter\ceil{\lceil}{\rceil}
\DeclarePairedDelimiter\floor{\lfloor}{\rfloor}

\usepackage{dsfont}
\usepackage{lettrine} % \lettrine[findent=1pt]{{{R}}}{}
\usepackage{amsmath,epsfig,amssymb,algpseudocode,amsthm,cite,url}
\usepackage{caption}
\allowdisplaybreaks
\usepackage{csquotes}
\usepackage{multirow}

\DeclareMathOperator*{\argmin}{arg\,min}
\usepackage{verbatim}
\usepackage{subfig}
\usepackage[english]{babel}
\usepackage{amsmath,amssymb}

\usepackage{verbatim}

\newtheorem{corollary}{Corollary}
\newtheorem{theorem}{\bf Theorem}

\newtheorem{lemma}{\bf Lemma}

\begin{document}
		\setlength{\abovedisplayskip}{3pt}
	\setlength{\belowdisplayskip}{3pt}
	\setlength{\abovedisplayshortskip}{3pt}
	\setlength{\belowdisplayshortskip}{3pt}
\title{\LARGE Wireless-Enabled Asynchronous Federated Fourier Neural Network for Turbulence Prediction in Urban Air Mobility (UAM)
}

\author{\IEEEauthorblockN{Tengchan Zeng, \emph{Student Member}, \emph{IEEE}, Omid Semiari, \emph{Member}, \emph{IEEE},\\  Walid Saad, \emph{Fellow}, \emph{IEEE}, and Mehdi Bennis, \emph{Fellow}, \emph{IEEE}\vspace{-0.35in}}\\
%	\IEEEauthorblockA{
%		\small $^1$Wireless@VT, Electrical and Computer Engineering Department, Virginia Tech, Blacksburg, VA, 24061 USA,\\ Emails:\url{{tengchan , walids}@vt.edu}.\\
%		$^2$Department of Electrical and Computer Engineering, University of Colorado Colorado Springs,\\Colorado Springs, CO,  80918 USA, Email:\url{osemiari@uccs.edu}.\\
%		$^3$Department of Electrical Engineering, Princeton University,
%		Princeton, NJ, 08544 USA,\\ Email:\url{mingzhec@princeton.edu}.\\		
%		$^4$Centre for Wireless Communications, University of Oulu, Oulu, 90014 Finland, Email:
%		\url{bennis@ee.oulu.fi}\vspace{-0.2cm}.
		\thanks{
		A preliminary version was presented at the IEEE Global Communications Conference, 2021 \cite{icc2021Zeng}. 
		This research was supported by the Office of Naval Research (ONR) under MURI Grant N00014-19-1-2621, the U.S. National Science Foundation under Grants CNS-1941348, and CNS-2008646, and by the Academy of Finland Project CARMA, by the Academy of Finland Project MISSION, by the Academy of Finland Project SMARTER, as well as by the INFOTECH Project NOOR.
		
       	T. Zeng and W. Saad are with Wireless@VT, Department of Electrical and
       	Computer Engineering, Virginia Tech, Blacksburg, VA, 24061 USA. E-mail:
       	\{tengchan, walids\}@vt.edu.
       	
       	O. Semiari is with the Department of Electrical and Computer Engineering,
       	University of Colorado, Colorado Springs, CO, 80918 USA. 
 	    E-mail: osemiari@uccs.edu.
 	    
 	    M. Bennis is with the Centre for Wireless Communications, University of Oulu, 90014 Oulu, Finland. E-mail:mehdi.bennis@oulu.fi.
}}
\maketitle

\begin{abstract}
	To meet the growing mobility needs in intra-city transportation, the concept of urban air mobility (UAM) has been proposed in which vertical takeoff and landing (VTOL) aircraft are used to provide a ride-hailing service. 
	In UAM, aircraft can operate in designated air spaces known as \emph{corridors}, that link the aerodromes, thus avoiding the use of complex routing strategies such as those of modern-day helicopters and alleviating the burden on the ground transportation system.  
	%A reliable communication network between GBSs and aircraft enables UAM to adequately utilize the airspace and create a fast, efficient, and safe transportation system. 
	For safety, a UAM aircraft must use air-to-ground communications to report flight plan, off-nominal events, and real-time movement to ground base stations (GBSs). A reliable communication network between GBSs and aircraft enables UAM to adequately utilize the airspace and create a fast, efficient, and safe transportation system. 
	In this paper, to characterize the wireless connectivity performance for UAM, a suitable spatial model is proposed. 
	%In particular, the distribution of GBSs is modeled as a Poisson point process (PPP), and the aircraft are distributed according to a combination of PPP, Poisson cluster process (PCP), and Poisson line process (PLP).
	For the considered setup, assuming that any given aircraft communicates with the closest GBS, the distribution of the distance between an arbitrarily selected GBS and its associated aircraft and the Laplace transform of the interference experienced by the GBS are derived. 
	Using these results, the signal-to-interference ratio (SIR)-based connectivity probability is determined to capture the connectivity performance of the UAM aircraft-to-ground communication network.
	Then, leveraging these connectivity results, a wireless-enabled asynchronous federated learning (AFL) framework that uses a Fourier neural network is proposed to tackle the challenging problem of turbulence prediction during UAM operations. 
	For this AFL scheme, a staleness-aware global aggregation scheme is introduced to expedite the convergence to the optimal turbulence prediction model used by UAM aircraft. 
	Simulation results validate the theoretical derivations for the UAM wireless connectivity. 
	The results also demonstrate that the proposed AFL framework converges to the optimal turbulence prediction model faster than the synchronous federated learning baselines and a staleness-free AFL approach. 
	Furthermore, the results
	characterize the performance of wireless connectivity and convergence of the aircraft's turbulence model under different parameter settings, offering useful UAM design guidelines. 
	% provide useful UAM design guidelines by showing the wireless connectivity and turbulence prediction model convergence performance under different parameter settings. 
\end{abstract} 
\section{Introduction}
According to the world urbanization prospects released by the United Nations, by 2030, more than 60\% of the world's population will live in urban areas and this percentage will jump to 70\% by 2050 \cite{unitednation}. 
Given this growth, mobility demands will push the ground transportation system to its limits, leading to a long commute for the public and significant economic costs for the society. 
In order to meet future mobility needs, the novel concept of Urban air mobility (UAM) was proposed \cite{FAA}. 
UAM will introduce vertical takeoff and landing (VTOL) aircraft to integrate the third dimension, i.e., airspace above cities, into the urban transportation system.  
%is proposed where fully automated vertical take off and landing (VTOL) aircraft is used as on-demand service, offering the third dimension, i.e., airspace above cities, into the urban transportation system. 
Recent technology advances on distributed electric propulsion, electrical energy storage, lightweight airframe structures, and electric VTOL are rapidly making UAM a reality \cite{FAA}. 
\subsection{UAM System Overview}
\begin{figure}[!t]
	\centering
	\includegraphics[width=4.2in,height=2.1in]{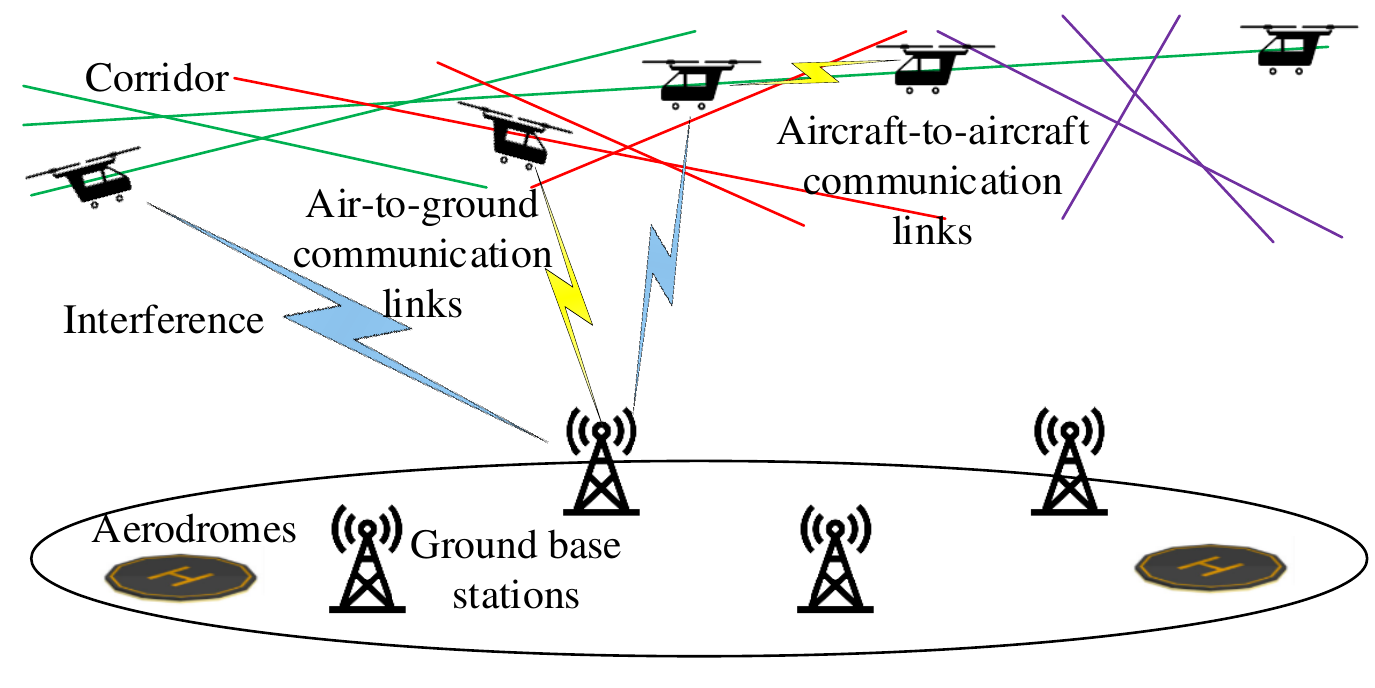}
	\DeclareGraphicsExtensions.
	\caption{Illustration of the UAM system which is composed of VTOL aircraft, aerodromes, corridors, and GBSs. To secure a safe operation, the UAM aircraft will constantly communicate with the GBS to submit the flight plan and convey off-nominal events.}
	\label{system_model1}
	%\vspace{-0.1in}
\end{figure}

As shown in Fig. \ref{system_model1}, a UAM system is composed of VTOL aircraft, aerodromes, corridors, and ground base stations (GBSs).
In particular, aerodromes are designed to support the arrival and departure operation of the aircraft. 
Moreover, the corridors linking different aerodromes constitute the airspace designated for UAM operations. 
Different from complex routing strategies currently used for helicopter applications, the idea of corridors can dramatically reduce the complexity of operations \cite{FAA}. 
Also, due to common mobility patterns shared by the public, corridors are usually concentrated around centralized points (CPs), such as residential, shopping, and business areas. 
In addition, GBSs in UAM will function as service providers that constantly communicate with the aircraft and deliver necessary information (e.g., weather and terrain), approve the flight plan submitted by the aircraft, and monitor the aircraft movements.

Each UAM operation has two major phases: \emph{planning} and \emph{en-route}. 
In the planning phase, once an aircraft receives a travel request between two aerodromes from an individual customer, it will determine the flight plan (e.g., the route selection of corridors, estimated travel time, and the aerodromes) which is subsequently submitted to a GBS via an air-to-ground link. 
Then, the GBS will evaluate the submitted plan against pre-determined constraints, such as the availability of corridors and aerodromes as well as possible conflicts with on-going operation of other aircraft. 
If the flight plan meets these pre-determined constraints, then the GBS will approve the flight plan and share it with other GBSs via backhaul links.
The aircraft will later operate in an en-route phase to pick up customers, navigate along the selected corridors, and arrive at the destination aerodrome within an anticipated travel time.
Note that, during the en-route phase, the aircraft must constantly communicate with the GBSs to convey any off-nominal events (e.g., deviation from the selected corridors) due to high winds and navigation degradation, as well as to accordingly update its flight plan.
%Meanwhile, the GBS will share the approved flight plan with other GBSs via backhaul. 
%Meanwhile, the aircraft will also communicate with surrounding aircraft to share the operational intent information so as to achieve the safe separation. 
%Hence, reliable air-to-ground communications are of great importance for the effective and safe operation of UAM. 
%
%Since UAM operation is still at its infancy, most of the related research focus on market studies \cite{goyal2018urban}, public acceptance \cite{al2020factors}, and operational
%constraints \cite{vascik2020geometric}. 
%There are few works studying the technical aspects of UAM operation. 
%For example, a learning-based collision avoidance is proposed in \cite{rodionova2020learning} to allow multiple aircraft operating independently and keeping a safe distance among each other. 
%In \cite{pradeep2019energy}, the arrival process of aircraft is optimized to minimize the energy consumption while meeting the expected travel time. 
%It is clear that, despite of the important role of communication networks in UAM operation, there is a lack of study to analyze the wireless connectivity of the aircraft when using UAM. 

\subsection{Motivation and Related Works}
Since UAM operation is still at its infancy, most of the related research focus on market studies \cite{goyal2018urban}, public acceptance \cite{al2020factors}, and operational constraints \cite{vascik2020geometric}. There are few works studying the technical aspects of UAM operation. 
For example, a learning-based collision avoidance is proposed in \cite{rodionova2020learning} to allow multiple aircraft operating independently and keeping a safe distance among each other. In \cite{pradeep2019energy}, the arrival process of
aircraft is optimized to minimize the energy consumption while meeting the expected travel time. 
Moreover, the air traffic management is studied in \cite{cotton2018airborne} and \cite{yang2019multi} to navigate the UAM aircraft through obstacles in a congested
urban area. 
In addition, the authors in \cite{donateo2020modeling} propose a battery model to quantify the effect of battery aging on the performance of electric flight in UAM.  
It is clear that, despite of the important role of communication networks in UAM operation, there is a lack of a rigorous, holistic analysis of the wireless connectivity of aircraft-to-GBS communication network in UAM. 

Beyond analyzing the connectivity needs of aircraft, another challenge in the UAM operation will be dealing with turbulence, i.e., air flow velocity changes around the aircraft \cite{anderson2010fundamentals}.
%.  
%In particular, given that UAM operation is still at its infancy, most of the related research focus on market studies \cite{goyal2018urban}, public acceptance \cite{al2020factors}, and operational constraints \cite{vascik2020geometric}. 
%There are only few works studying the technical aspects of UAM operation. 
%For example, a learning-based collision avoidance is proposed in \cite{rodionova2020learning} to allow multiple aircraft to operate independently and maintain a safe distance. 
%In \cite{pradeep2019energy}, the arrival process of aircraft is optimized to minimize the energy consumption while meeting the estimated travel time.
%The authors in \cite{8569645} propose a travel time estimation scheme for UAM aircraft that takes into account several constraints such as battery limitations, corridor occupancy, and aerodrome capacity.
%Other than the collision avoidance, trajectory design, and travel time estimation, another major challenge in UAM operation is the turbulence, i.e., chaotic changes in air flow velocity.
Different from commercial jet aircraft, UAM aircraft are more likely to be affected by turbulence due to the following two facts.
First, UAM aircraft can only operate at the troposphere \cite{FAA} where the aerodynamics are constantly changing because of varying terrains, altitudes, and temperatures, in contrast to the commercial jets which usually cruise in the stable stratosphere. 
Second, compared to commercial jets, the lighter weight and smaller size of UAM aircraft will increase their susceptibility to turbulence, imposing a huge safety risk for passengers. 
However, if the air flow velocity can be predicted over time, UAM aircraft can take proper actions (i.e., yaw, pitch, and roll) to adjust their rotations and movements so as to minimize the impact of turbulence on the overall operation, avoiding the deviation from corridors and securing the safety. 
Hence, an accurate turbulence prediction will be a key enabler for the UAM operation. 
%In particular, with the accurate turbulence prediction, the UAM aircraft can choose the operation decision to secure the basic safety. 
%As a result, the turbulence prediction can also improve the energy efficiency of VTOL aircraft which is another key challenges in the development of UAM system. 

%Since the air turbulence is usually modeled as Naiver-Stokers equations, one type of partial differential equations, which governs the internal and external force on the air particles, an accurate prediction will rely on solving these equations. 
To perform turbulence prediction, 
there is an increased interest in studying physical informed neural
networks where the physical knowledge is integrated into the neural network design.
%the most common way is to solve the PDEs, such as Burgers' equation \cite{bonkile2018systematic} and Navier-Stokes equation \cite{alfonsi2009reynolds},  which govern the internal and external forces among the air particles and capture the turbulence around UAM aircraft over time. 
%Unfortunately, solving these PDEs can be extremely challenging and time-consuming. 
%For instance, solving the Navier-Stokes equation is considered to be one of the seven millennium prize problems \cite{jaffe2006millennium}.  
%Instead, people use approaches, e.g., Reynolds-averaged Navier-Stokes (RANS) models \cite{alfonsi2009reynolds} and large eddy simulation (LES) \cite{sagaut2006large}, to obtain approximated solutions where the accuracy is usually limited and the turbulence predictions can be highly unreliable. 
%To achieve accurate predictions without directly solving these PDEs, there is an increased interest in studying physical informed neural networks where the physical knowledge is integrated into the neural network design \cite{10.1145/3394486.3403198,kochkov2021machine,li2020fourier}. 
For example, the authors in \cite{10.1145/3394486.3403198} and \cite{kochkov2021machine} use neural networks to build a mapping between the input data (e.g., temperature and pressure) and turbulence predictions in the Euclidean space. 
In contrast, the Fourier neural network (FNN), proposed in \cite{li2020fourier}, considers data in the Fourier space. 
The motivation behind the use of FNN is that fluid dynamics (e.g., turbulence flow) are usually composed of components at different frequencies where the general direction is controlled by components at low frequency and the little eddies in the fluid flow are consisted of high-frequency components. 
By considering the input and output in the Fourier space, FNN can achieve a better accuracy than other learning based turbulence prediction models \cite{li2020fourier}. 
However, when using the learning methods proposed in  \cite{10.1145/3394486.3403198,kochkov2021machine,li2020fourier} for turbulence prediction, the local data can be insufficient to train the learning model due to the limited on-chip memory available on board UAM aircraft. 
%ignore the insufficient and possibly skewed local training data led by the limited storage at CAVs. 
%When using physical informal neural networks to solve PDEs for turbulence predictions , an individual aircraft can only store data related to its most recent operations due to the limited storage. 
%However, such data can be easily skewed and of poor quality. 
As a result, when the UAM aircraft operate in a new environment or encounters less frequently occurred events (e.g., poor weather), the turbulence models in \cite{10.1145/3394486.3403198,kochkov2021machine,li2020fourier} can fail to adapt to such changes, jeopardizing the operation safety. 
Therefore, an effective turbulence predictor design will directly depend on \emph{collaboratively and jointly} training the prediction model among multiple UAM aircraft. 
%In particular, since the turbulence flow is spatially and temporally correlated \cite{WALLACE2014022003}, such a collaboratively learned turbulence prediction model can enable a particular aircraft to adapt to new turbulence dynamics unknown to the aircraft but experienced by other aircraft.
To this end, one can use federated learning (FL) \cite{chen2020wireless} to build a cooperative learning framework in which a centralized unit, such as the GBS, aggregates the knowledge learned by a group of UAM aircraft.  

%By using FL, UAM aircraft can train the turbulence prediction model based on their own data, and, then, a parameter server (i.e., the GBS) can consistently aggregate the trained model parameters received from the aircraft.  
However, it will be challenging to use conventional FL frameworks \cite{konevcny2016federated,li2020federated,bonawitz2019towards,reisizadeh2020fedpaq,zeng2021federated,8945292,8843942,xie2019asynchronous,nguyen2021federated}
% FedAvg \cite{konevcny2016federated} and FedProx \cite{li2018federated}, 
for UAM aircraft turbulence prediction.
On the one hand, synchronous FL algorithms, such as those in \cite{konevcny2016federated,li2020federated,bonawitz2019towards,reisizadeh2020fedpaq,zeng2021federated}, are not suitable
for UAM aircraft turbulence prediction due to the following two reasons.  
First, for synchronous FL frameworks, like FedAvg \cite{konevcny2016federated} and FedProx \cite{li2020federated}, 
%the new communication round will not start unless all local users complete the model training and transmit the model parameters to the parameter server. 
UAM aircraft will suffer from the straggler effect where some local aircraft take much longer time than others to send their learned knowledge to the GBS, thereby increasing the convergence time and jeopardizing the aircraft's ability to quickly predict the turbulence. 
Second, for synchronous FL schemes, like scalable FL \cite{bonawitz2019towards}, FedPAQ \cite{reisizadeh2020fedpaq}, and DFP \cite{zeng2021federated}, 
%the parameter server will update the global model by solely considering the model parameters from a certain number of local aircraft which can complete model training and transmission faster than others.   
%When using such synchronous FL in UAM, 
not all UAM aircraft can participate in the FL training and some important local model updates can be discarded, leading to a poor real-time turbulence prediction performance. 
On the other hand, since the staleness associated with the local trained model parameters is not properly considered, the prior work on asynchronous FL (AFL) \cite{8945292,8843942,xie2019asynchronous,nguyen2021federated} is not suitable for handling the UAM turbulence prediction task.
For instance, these prior works either ignore staleness (e.g., \cite{8945292} and \cite{8843942}) or consider impractical assumptions, like bounded staleness \cite{xie2019asynchronous} and \cite{nguyen2021federated}, in their designed frameworks.
However, 
because of the varying wireless channel conditions and mobility of aircraft, the staleness associated with the local model update will be unbounded and also vary from one UAM aircraft to another. 
If not being properly considered in the AFL framework design, such a unbounded and randomly distributed staleness can impede the convergence to the optimal turbulence prediction model used by UAM aircraft \cite{dai2018toward}.
%However, these prior art either ignore the impact of staleness on the convergence (e.g., \cite{8945292}) or consider impractical assumptions, like bounded staleness \cite{xie2019asynchronous}, in the convergence study. 
%These AFL framework will not fit the turbulence prediction among UAM aircraft. 
%However, due to the large distance between the UAM aircraft and GBS and obstruction
%by other vehicles, buildings, trees, pedestrian, or other objects, it is highly probable that only non-line-of-sight (NLoS) links between aircraft to GBS exist.
%Along with fading channels in the UAM communication network, the NLoS link will lead to  randomly distributed and unbounded staleness among local aircraft. 
%Hence, in order to consider the staleness associated to local updates in our AFL framework, we must first conduct a performance analysis for UAM air-to-ground communication networks and determine the distribution of staleness. 
%Meanwhile, the AFL framework must be designed to mitigate the negative impact of staleness on the turbulence prediction model convergence at UAM aircraft. 

\subsection{Contributions and Outcomes}
The main contributions of this paper is a wireless-enabled AFL framework in which UAM aircraft use the communication network to collaboratively train a federated FNN and perform efficient turbulence prediction. 
In particular, to characterize how aircraft leverage wireless connectivity for AFL, we propose a stochastic geometry based spatial model for GBSs, corridors, and aircraft in UAM system. 
Moreover, we perform the wireless connectivity analysis for aircraft-to-GBS communication network in UAM.
%analyze the connectivity performance for UAM wireless network. 
Based on the wireless connectivity study, we propose a wireless-enabled AFL framework in which UAM aircraft use the wireless network to collaboratively optimize the FNN-based turbulence prediction model. 
To mitigate the impact of staleness on convergence, we study a staleness-aware AFL framework and analyze the convergence. 
The novelty of this work lies in the following key contributions:
%One contribution of this paper is a novel, rigorous wireless connectivity  performance analysis of aircraft-to-GBS communication network in UAM. 
%a wireless-enabled AFL framework in which UAM aircraft use the communication network to collaboratively perform the turbulence prediction. 
%In particular, to characterize how aircraft leverage wireless connectivity for AFL, we propose a stochastic geometry based spatial model for GBSs, corridors, and aircraft in UAM system. 
%Moreover, we analyze the connectivity performance for UAM wireless network. 
%%perform a connectivity performance analysis for the UAM wireless network and determine the number of UAM aircraft that will participate in the AFL as well as the distribution of staleness experienced by the local learning model updates. 
%Another contribution is that, based on the wireless connectivity study, we propose a wireless-enabled AFL framework in which UAM aircraft use the wireless network to collaboratively optimize the turbulence prediction model. 
%To mitigate the impact of staleness on convergence, we study a staleness-aware AFL framework and analyze the convergence. 
%%The novelty of this work lies in the following key contributions:
\begin{itemize}
	\item We perform a novel, rigorous performance analysis of aircraft-to-GBS communication networks in UAM. 
	In particular, by leveraging Poisson point process (PPP),
	Poisson cluster process (PCP), and Poisson line process (PLP) from stochastic geometry, we model the spatial distribution of GBSs, corridors, and aircraft
	in UAM.  
	Assuming that any given aircraft will communicate with the closest GBS, we further derive the Laplace transform of the interference experienced by a GBS and the communication distance distribution between the GBS and its associated aircraft.
	Using these results, we obtain the signal-to-interference ratio (SIR)-based connectivity probability for the aircraft-to-GBS communication network in UAM.
	\item We propose a wireless-enabled collaborative learning framework consisting of FNN and AFL to optimize the turbulence prediction model for UAM aircraft. 
	In particular, the local UAM aircraft will train an FNN model using their own data and, then, transmit the trained FNN parameters to the GBS via the UAM wireless network.  Next, the GBS will aggregate the received parameters and generate a new global model in an asynchronous fashion. 
	Based on the wireless connectivity study, we derive the staleness distribution of local FNN updates received at the GBS and the number of UAM aircraft that participate in AFL. 
	%To capture how UAM aircraft utilize wireless links for collaborative learning, we develop a stochastic geometry based spatial model that characterizes the distribution of GBSs, corridors, aircraft in UAM system using PPP, Poisson cluster process (PCP), and Poisson line process (PLP).
	\item To mitigate the impact of staleness on the overall convergence, we propose a staleness-aware global aggregation scheme for performing AFL.
	In particular, unlike conventional FL frameworks \cite{konevcny2016federated,li2020federated,bonawitz2019towards,reisizadeh2020fedpaq,zeng2021federated,8945292,8843942,xie2019asynchronous,nguyen2021federated}, the proposed staleness-aware global aggregation scheme explicitly considers the unbounded and randomly distributed staleness of locally trained FNN parameters. 
	We then analyze the convergence rate of the proposed AFL framework and highlight the importance of considering the staleness in the global aggregation.  
\end{itemize}

Simulation results validate our theoretical analysis for the wireless connectivity study.
Moreover, the results show that our AFL framework can achieve a faster convergence than the conventional synchronous FL counterparts and AFL framework without considering the staleness of local FNN parameters. 
In addition, the results show the wireless connectivity and FNN model convergence for different UAM parameter settings, offering useful system design insight for deploying UAM.
\emph{To the best of our knowledge, this is the first work that analyzes the connectivity performance of a UAM communication network and develops a staleness-aware AFL framework to optimize the turbulence prediction model for UAM aircraft.}
%Based on the performance analysis of the UAM communication network, a novel AFL framework is proposed to enable aircraft to collaboratively learn and optimize the FNN-based turbulence prediction model. 
%In particular, unlike the synchronous FL, our asynchronous approach can eliminate the straggler effect and aggregate the knowledge learned by all UAM aircraft.
%Moreover, the proposed AFL framework explicitly takes into account the staleness of the  local model updates received for the global aggregation. 
%We perform a detailed convergence study for the proposed AFL framework where we identify how the staleness affects the convergence of the optimal turbulence prediction model. 
%
%\emph{To the best of our knowledge, this is the first work that analyzes the connectivity performance of air-to-ground communication networks in a UAM scenario and studies a wireless-enabled AFL framework to optimize the turbulence prediction design for UAM aircraft.} 
%Simulation results are provided to validate our theoretical analysis for the connectivity study and show the performance for different parameters thus offering useful system design insight for deploying UAM. 

The rest of the work is organized as follows. Section \uppercase\expandafter{\romannumeral2}
presents the system model for UAM. 
Section \uppercase\expandafter{\romannumeral3}
provides a theoretical analysis of the aircraft's wireless connectivity.
Section \uppercase\expandafter{\romannumeral4} introduces the staleness-aware AFL framework and analyzes the convergence. 
Section \uppercase\expandafter{\romannumeral5} provides simulation results, and conclusions
are drawn in Section \uppercase\expandafter{\romannumeral6}.

\section{System model}
Consider a group of UAM aircraft, as shown in Fig. \ref{system_model1}.
During flight, each aircraft will share a flight plan
with other aircraft via aircraft-to-aircraft communications so as to maintain a safe distance and avoid collisions. 
Meanwhile, each aircraft constantly communicates with a GBS to report its location and possible off-nominal events, as well as transmit the turbulence prediction model updates.  
%The off-nominal events can be caused by the high winds, performance issues, and navigation degradation.  
%Meanwhile, each aircraft will also 
%share the flight plan with other aircraft via aircraft-to-aircraft communications so as to maintain a safe distance and avoid collisions during the operation.  
\subsection{Spatial Models for UAM}

\begin{figure}[!t]
	\centering
	\includegraphics[width=3.0in,height=2.6in]{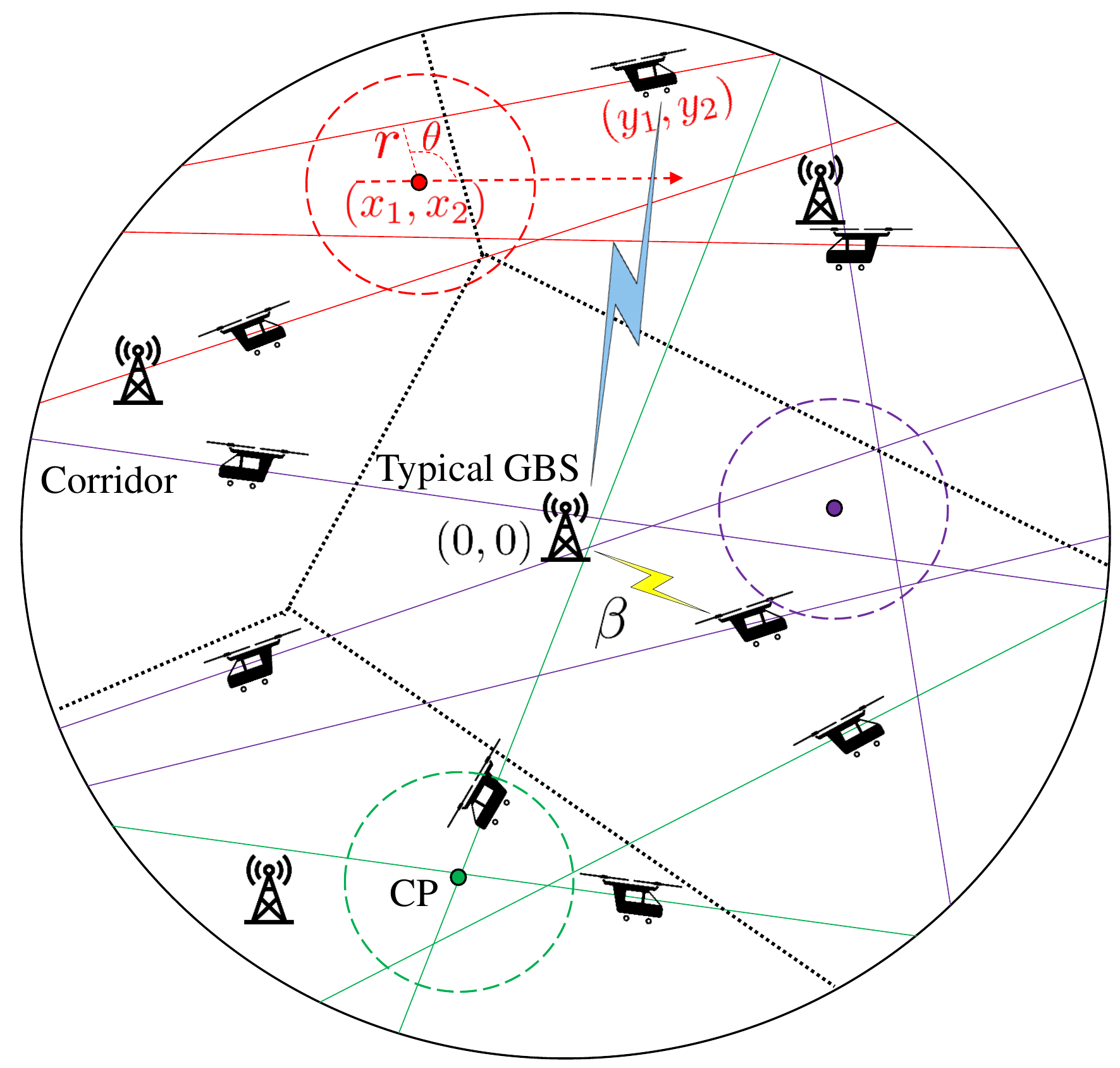}
	\DeclareGraphicsExtensions.	
	\caption{The top view of how corridors, aircraft, centralized points, and GBSs are distributed in UAM.}
	\label{system_model2}
	%\vspace{-0.1in}
\end{figure}
%\begin{figure}[!t]
%	\centering \vspace{-0.05in}
%	\subfloat[]{%
%		\includegraphics[width=2.5in,height=1.3in]{system_model_1.pdf}\label{system_model1}
%	}	
%
%	\subfloat[]{%
%		\includegraphics[width=2.0in,height=1.5in]{system_model_2.pdf}\label{system_model2}
%	}	
%	\caption{\small Illustration of urban air mobility (UAM) model. UAM is presented in (a) where the base station receives the uplink transmission coupled with interference. The top view of how corridors, aircraft, centralized points, and base stations are distributed in UAM is shown in (b).\vspace{-0.25in}}
%	\label{system_model}	
%\end{figure}
To characterize relative locations of different components in UAM, we must study the spatial modeling of aircraft and GBSs. 
As shown in Fig. \ref{system_model2}, we model the distribution of GBSs as a two-dimensional PPP $\Xi$ with density $\lambda_{b}$. 
For the distribution of aircraft, we first model the spatial distribution of CPs and corridors. 
In particular, to capture the fact that CPs can spread across space, we model their distribution as a two-dimensional PPP $\Phi$ with density $\lambda_{c}$. 
Around CP $i \in \Phi$, we model the distribution of corridors as a PLP $\Psi_{i}$ with density $\lambda_{l}$ where all corridors share the same height $h$.
Since there will be a limited number of corridors around each CP, we  assume that the maximum number of corridors for each CP is $N$.
In this case, the number of corridors follows a truncated Poisson distribution within the range $[0,N]$.

Similar to \cite{chiu2013stochastic}, the relative location of a corridor around a CP is determined by two factors: the distance $r \in \mathbb{R}_{+}$ between the CP and the corridor and the angle $\theta \in [0,2\pi)$ between the positive $x$-axis and the line  perpendicular to the corridor, as shown in Fig. \ref{system_model2}.
However, instead of using a randomly distributed $r$ in $\mathbb{R}_{+}$ that would not capture the realistic spatial location of centralized corridors, inspired by PCP \cite{haenggi2012stochastic}, we model the distance $r$ using the following two approaches.
The first approach is aligned with the Thomas cluster process whereby we model the distance $r$ as a truncated Gaussian distribution as follows: $f_{R}(r)\!=\!\sqrt{\frac{2}{\pi \sigma^{2}}}\exp\!{\left(-\frac{r^2}{2\sigma^2}\right)},  r \in [0,\infty).$ 
In the second approach, we model the distance $r$ as a uniform random variable within $[0,\hat{r}]$ with $\hat{r}$ being the maximum distance from the CP, i.e., $f_{R}(r)=\frac{1}{\hat{r}}, r \in [0,\hat{r}].$
As such, we can guarantee that all corridors around the same CP will pass through a circular disc area with radius $\hat{r}$, similar to the definition of the Matern cluster process.
With these two approaches, we can characterize the fact that the corridors are closely distributed around their CPs instead of being spread over the whole space as considered in \cite{chiu2013stochastic}. 
Based on the real deployment and travel demands from the public, we can use either one of these two approaches to model the distance $r$.
Next, on a randomly selected corridor $j \in \Psi_{i}$, we assume that the distribution of transmitting aircraft follows a one-dimensional PPP with density $\lambda_{t}$.
%If the probability that each aircraft functions as the transmitter is $\rho$, then the locations of the transmitting aircraft on corridor $j \in \Psi_{i}$ can be modeled as a one-dimensional PPP $\Omega_{i,j}$ with density $\lambda_{t} = \rho\lambda_{a}$ according to the PPP thinning theorem \cite{chiu2013stochastic}.  
\subsection{Aircraft-to-GBS Communication Model}
Similar to prior works \cite{8340239,8419219,andrews2016primer}, we assume that each aircraft will be served by the closest GBS. 
Due to the stationarity of the two-dimensional PPP, we arbitrarily select a GBS as the \emph{typical GBS} and assume that it is located at the origin with zero height. 
Thus, we can calculate the received signal power at the typical GBS from its associated aircraft as
\begin{align}
	\label{(1)}
	P 	= p g (h^2 + \beta^2)^{-\frac{\alpha}{2}}, 
\end{align} 
where $p$ is the transmit power used by all aircraft, and $\beta$ is the distance between the vertical projection of the associated GBS at the plane with height $h$ and the associated aircraft. 
$\alpha$ is the path loss exponent, and $g$ is the air-to-ground wireless channel gain.
We model the communication channels as independent Nakagami channels with an integer $m$ to characterize a wide range of fading environment.

While receiving the transmission from the associated aircraft, the typical GBS will experience interference from two sources. 
The first one relates to aircraft who communicate with other GBSs rather than the typical GBS. 
The second interference source is aircraft who share the flight plan with surrounding aircraft. 
Taking into account the distribution of the transmitter aircraft in UAM, we can calculate the interference at the typical GBS as follows 
\begin{align}
	\label{(2)}
	I = \sum_{i\in\Phi} \sum_{j\in \Psi_{i}} \sum_{k \in \Omega_{i,j}} p g' (||\boldsymbol{x}+\boldsymbol{y}||^2+h^2)^{-\frac{\alpha}{2}},
\end{align} 
where, as shown in Fig. \ref{system_model2}, $\boldsymbol{x}=(x_{1},x_{2})$ is the  location of CP $i\in\Phi$ relative to the vertical projection of the typical GBS at the plane of height $h$, and $\boldsymbol{y}=(y_{1},y_{2})$ denotes the relative location of an interfering aircraft compared to its CP.

%In this case, the signal-to-noise-plus-interference (SINR) can be approximated as SIR $V$.

\subsection{Learning Model}
\begin{figure}[!t]
	\raggedleft
	\includegraphics[width=6.5in,height=1.8in]{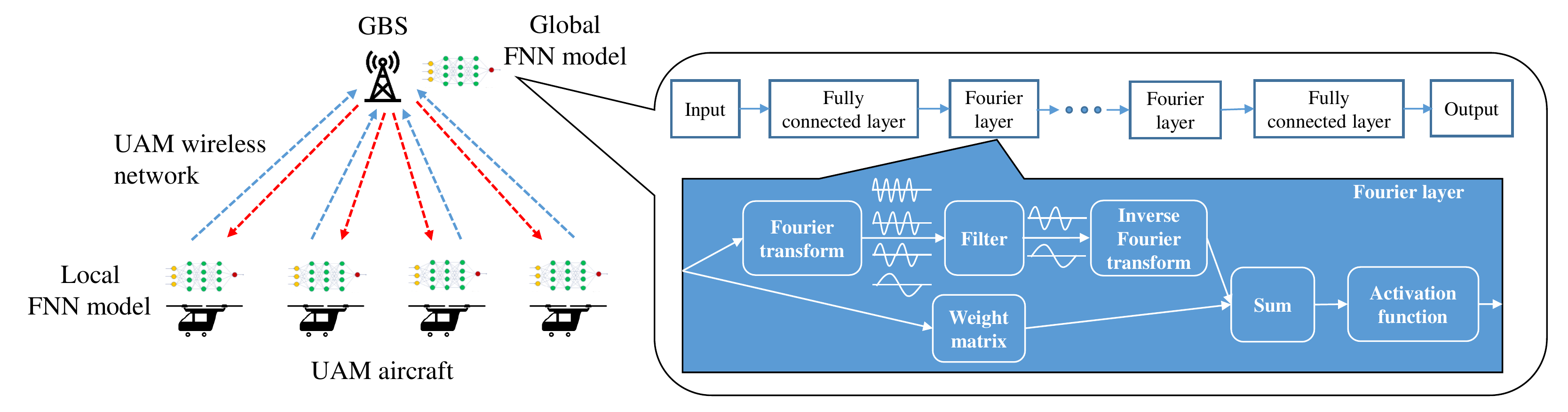}
	\DeclareGraphicsExtensions.
	\caption{UAM aircraft use a combination of FNN and AFL to optimize the turbulence prediction.}
	\label{FNN}
	%\vspace{-0.1in}
\end{figure}

%Navier–Stokes equations are usually used to model and predict the turbulence around the aircraft, and such partial differential equations are 
%\begin{align}
%	\label{weight}
%	&\nabla \boldsymbol{v} = 0, \\  
%	\label{force}
%	& \frac{\partial \boldsymbol{v}}{\partial t} + (\boldsymbol{v}) \cdot \nabla \boldsymbol{v}= \frac{-1}{\rho_{0}}\nabla \varrho + \nu \nabla^2\boldsymbol{v} + f, 
%\end{align}
%where $\boldsymbol{v}$ is the velocity field of the air flow around the aircraft with components in the three-dimensional space. $\varrho$ and $\rho_{0}$ are, respectively, the air pressure and density, and $\varrho$ is the kinematic viscosity, $f$ is the body force led by the earth gravity. 
%The Navier–Stokes equations in (\ref{weight}) and (\ref{force}) in essence express the conversations of mass and of momentum for air. 
%Solving the Navier–Stokes equations is notoriously difficult and it is one of seven Millennium Prize Problems \cite{jaffe2006millennium}.
%As a result, people tend to rely on approaches, such as Reynolds averaging method, to obtain the approximated solution where the accuracy is hardly guaranteed.  
To accurately predict turbulence, the UAM aircraft will use a combination of FNN and asynchronous FL. 
In particular, the aircraft will use the architecture of FNN \cite{li2020fourier} to train their local data, as shown in Fig. \ref{FNN}.
The input and output for the FNN will be, respectively, the turbulence flow history data and prediction. 
Different from conventional machine learning methods that build an approximated function between the inputs and outputs defined in the Euclidean space, the FNN consists of Fourier layers whereby the data in the Fourier space is explicitly considered. 
The reason for considering data in the Fourier space is due to the fact that the turbulence flow can be decomposed into components at different frequencies. 
As shown in Fig. \ref{FNN}, by filtering out the components at high frequencies in the Fourier layers, the FNN can reduce the impact of high-frequency noise existing in the data collected at the aircraft, and, thereby, increase the turbulence prediction accuracy. 

In AFL, the GBS will act as the parameter server and the set $\mathcal{K}$ of $K$ associated aircraft will collaboratively learn the FNN-based turbulence prediction model. 
In particular, the GBS will first generate an initial global FNN parameters $\boldsymbol{w}_{0}$ for the FNN model and broadcast it to all associated aircraft. 
Then, the aircraft will use the received FNN parameters to train its local data and transmit the updated FNN parameters back to the associated GBS in the uplink. 
Next, whenever receiving the local FNN parameter updates from any arbitrary aircraft, the GBS will update the global FNN parameters and share the updated parameters with the corresponding aircraft. 
This AFL process is repeated over uplink-downlink channels between the GBS and the associated aircraft, and the global and local FNN parameters will be sequentially updated. 
As communication rounds proceed, the model will converge to the optimal turbulence prediction model which is the solution to the following optimization problem: 
\begin{align}
	\label{optimal}
	&\argmin_{\boldsymbol{w}} \sum_{k=1}^{K}\sum_{j=1}^{s_{k}}\frac{s_{k}}{s_{K}}f_{k}(\boldsymbol{w},\xi_{j}), \\
	&\hspace{0.04in}\text{s.t.} \hspace{0.1in} \boldsymbol{w}^{(1)} = \boldsymbol{w}^{(2)} = ... = \boldsymbol{w}^{(K)} = \boldsymbol{w},
\end{align}
where $s_{K}=\sum_{k=1}^{K}s_{k}$ is the size of the training data across all aircraft with $s_{k}$ being the size of training data at aircraft $k$.
$f_{k}(\boldsymbol{w},\xi_{j})$ is the loss function of aircraft $k\in \mathcal{K}$ when using the FNN model parameter $\boldsymbol{w}$ to train local data $\xi_{j}$, and $f_{k}(\boldsymbol{w})=\sum_{j=1}^{s_{k}}f_{k}(\boldsymbol{w},\xi_{j})$ is the total loss at aircraft $k\in \mathcal{K}$.

Clearly, in the local FNN training phase, some aircraft use the most updated global model while others can only train stale versions of global model over their data.
Given the higher computing power and larger communication bandwidth of a GBS, the local FNN parameters can be updated immediately once the GBS receives the trained model updates from the corresponding aircraft. 
In this case, \textit{the staleness associated to local FNN parameters can thereby defined as the time
elapsed since the generation of the local model parameter update that is most recently received at the GBS.}
If the aircraft complete the local model training and model parameter transmission with less time, their trained FNN parameters will be associated with less staleness. 
Hence, staleness can be calculated as the time spent on the FNN model training and FNN parameters transmission. 
For the FNN model training, the computing delay can be derived as $t_{\text{comp}}=\frac{v\epsilon}{\varepsilon}$, where $v$ is size of training data, $\epsilon$ refers to the number of computing cycles needed per bit, and $\varepsilon$ is the frequency of CPU clock of UAM aircraft. 
When calculating the transmission delay, for tractability, we assume that the noise is negligible compared to the interference. 
%We also assume that the computing time is negligible in contrast to the communication delay and, thereby, the staleness of local is dominated by the transmission delay. 
Therefore, given the received signal and interference calculated in (\ref{(1)}) and (\ref{(2)}), the transmission delay will be as $t_{\text{tran}}= \frac{V}{W\log_{2}(1+\frac{P}{I})}$, where $V$ is the size of the data packet containing the FNN parameters and $W$ is the communication bandwidth. 
Hence, the staleness associated to the locally trained FNN parameters will be given by $\delta=t_{\text{comp}}+t_{\text{tran}}$.

When designing the AFL framework to optimize the turbulence prediction model for UAM aircraft, we need to address a number of challenges. 
First, to characterize the convergence of the AFL framework, we must determine the number of UAM aircraft that participate in AFL and the staleness distribution of local FNN updates. 
To this end, one can use tools from stochastic geometry to analyze the performance of a UAM communication network. 
However, existing stochastic geometry approaches, such as those in \cite{8340239,8419219,andrews2016primer},  cannot be directly applied to our model. 
This is because, different from conventional networks (e.g., cellular systems) that can be simply modeled by a single point process, 
the spatial distribution of UAM aircraft is characterized by a combination of PPP, PCP, and PLP. 
Meanwhile, due to the fading channels in the UAM wireless network and the mobility of UAM aircraft, the staleness of locally trained FNN parameters will be unbounded and randomly distributed.
If the AFL framework is not designed properly, such varying staleness can have a detrimental effect on the overall convergence. 
Hence, the second challenge will be determining and mitigating the impact of such unbounded and randomly distributed staleness on the overall FNN model convergence. 
In this way, UAM aircraft can quickly and accurately make the turbulence prediction and take proper actions to secure the operation safety. 

In Section \uppercase\expandafter{\romannumeral3}, we take into account the complex distribution of UAM aircraft and analyze the connectivity performance as well as derive the staleness distribution of local FNN parameters and the number of aircraft participating in the AFL.
Then, we propose a staleness-aware AFL framework and analyze its convergence in Section \uppercase\expandafter{\romannumeral4}.
%{\color{blue} More challenges associated to the FNN...}
\section{Performance Analysis of UAM Aircraft-to-GBS Communication Network} 
First, we characterize the distance distribution of the typical GBS and its associated aircraft. Then, we calculate the Laplace transform of the interference experienced by the typical GBS. 
Next, we derive the SIR-based connectivity probability for the aircraft when operating in the UAM corridors. 
Based on these results, we derive the distribution of staleness of local FNN parameters and the number of aircraft that participate in the AFL.

\subsection{Distance Distribution between the Typical GBS and its Associated Aircraft}
Since each aircraft communicates with its closest GBS, we can derive the statistical distribution of the distance $\beta$ between the vertical projection of the typical GBS and its associated aircraft in the following lemma that  follows from \cite{andrews2016primer}.
\begin{lemma}
	\label{lemma1}
	\emph{For a group of GBSs whose distribution follows a two-dimensional PPP with density $\lambda_{b}$, the probability density function (PDF) of the distance $\beta$ between the vertical projection of the typical GBS and its associated aircraft is $f_{B}(\beta) = 2\pi \lambda_{b} \beta \exp(-\lambda_{b}\pi \beta^2)$. }
\end{lemma}
With Lemma \ref{lemma1}, we can find the statistical distribution of the distance between the typical GBS and its associated aircraft at height $h$ and further use (\ref{(1)}) to determine the received signal power at the typical GBS. 

\subsection{Laplace Transform of Interference at the Typical GBS}
\begin{figure}[!t]
	\centering
	\includegraphics[width=4in,height=0.8in]{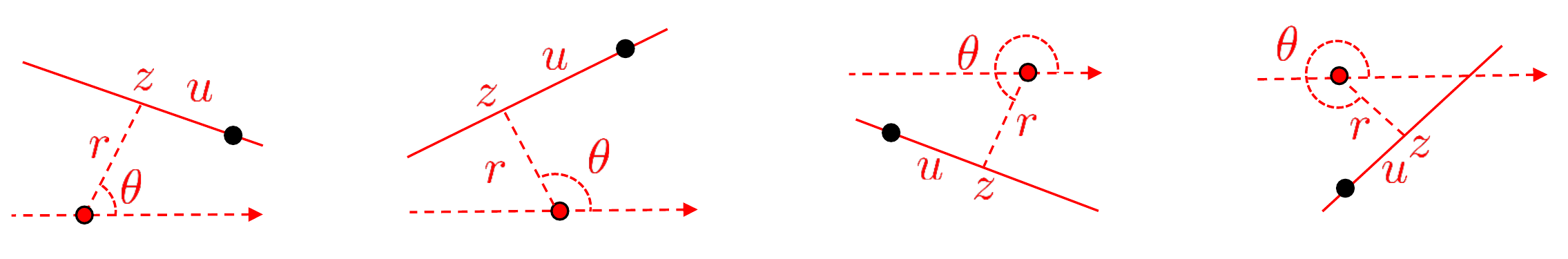}
	\DeclareGraphicsExtensions.
	\caption{Four possible relative locations between CP and corridors where the black solid point, red solid point, and solid line, respectively, show the aircraft, CP, and corridor. }
	\label{y_presentation}
	%\vspace{-0.1in}
\end{figure}

In order to capture the interference experienced by the typical GBS, we need to model the distance between the typical GBS and the interfering aircraft. 
To this end, we start with the relative location $\boldsymbol{y}=(y_{1},y_{2})$ of the interfering aircraft around its CP. 
As shown in Fig. \ref{y_presentation}, we list four possible relative locations between CP and corridors. 
Assume that the projected point of CP on the corridor is $\boldsymbol{z} = (z_{1},z_{2})$ and the distance between $\boldsymbol{z}$ and the aircraft is $u \in \mathbb{R}$. 
Hence, we can obtain the four alternatives for the location of aircraft on the corridor as 
\begin{equation*}
	\begin{cases}
		(z_{1}+u \cos(\frac{\pi}{2}-\theta),z_{2}-u \sin(\frac{\pi}{2}-\theta) ), \quad &\text{if} \, \theta \in [0,\frac{\pi}{2}), \\
		(z_{1}+u \cos(\theta-\frac{\pi}{2}),z_{2}+u \sin(\theta-\frac{\pi}{2}) ), \quad &\text{if} \, \theta \in [\frac{\pi}{2},\pi), \\
		(z_{1}-u \cos(\frac{3\pi}{2}-\theta),z_{2}+u \sin(\frac{3\pi}{2}-\theta) ), \quad &\text{if} \, \theta \in [\pi,\frac{3\pi}{2}), \\
		(z_{1}-u \cos(\theta-\frac{3\pi}{2}),z_{2}-u \sin(\theta-\frac{3\pi}{2}) ), \quad &\text{if} \, \theta \in [\frac{3\pi}{2},2\pi). \\
	\end{cases}
\end{equation*}
Meanwhile, the location of the CP can be given by:
\begin{equation*}
	\begin{cases}
		(z_{1}-r \cos(\theta),z_{2}-r \sin(\theta) ), \quad &\text{if} \, \theta \in [0,\frac{\pi}{2}), \\
		(z_{1}+r \sin(\theta-\frac{\pi}{2}),z_{2}-r \cos(\theta-\frac{\pi}{2}) ), \quad &\text{if} \, \theta \in [\frac{\pi}{2},\pi), \\
		(z_{1}+r \sin(\frac{3\pi}{2}-\theta),z_{2}+r \cos(\frac{3\pi}{2}-\theta) ), \quad &\text{if} \, \theta \in [\pi,\frac{3\pi}{2}), \\
		(z_{1}-r \cos(2\pi-\theta),z_{2}+r \sin(2\pi-\theta) ), \quad &\text{if} \, \theta \in [\frac{3\pi}{2},2\pi). \\
	\end{cases}
\end{equation*}

Given the locations of the aircraft and CP for $\theta \in [0,2\pi)$, we have $\boldsymbol{y}=(u \sin \theta+ r\cos \theta, -u \cos \theta + r\sin\theta)$, $r \in \mathbb{R}_{+}$ and $u \in \mathbb{R}$.
Therefore, the distance between the vertical projection point of the typical GBS and the interfering aircraft is $f(x_{1},x_{2},u,r,\theta)=||\boldsymbol{x}+\boldsymbol{y}||=(x_1+u \sin \theta+ r\cos \theta)^2+(x_{2}-u \cos \theta + r\sin\theta)^2)^{1/2}$.
Hence, we can determine the Laplace transform of the interference experienced by the typical GBS in the following lemma. 
\begin{lemma}
	\label{lemma2}
	\emph{When UAM aircraft operating within the corridors are distributed according to the one-dimensional PPP, the Laplace transform of the interference experienced by the typical GBS is}
	\begin{align}
		\label{(4)}
		\mathcal{L}(s)\!=\!\exp{\left(-\lambda_{c}\int_{\mathbb{R}^+}\int_{0}^{2\pi}\left(1-\sum_{n=0}^{N}\left(\mathcal{K}_{2}(l\cos\phi,l\sin \phi)\right)^n  \mathbb{P}(n|n<N)\right) ld\phi dl\right)},
	\end{align}
	\emph{where $\mathbb{P}(n|n\leq N) = \frac{(\lambda_{l}-1)^n e^{-(\lambda_{l}-1)}}{n!\omega}$ with $\omega= \sum_{n=0}^{N}\frac{(\lambda_{l}-1)^n e^{-(\lambda_{l}-1)}}{n!}$, and $\mathcal{K}_{2}(l\cos\phi,l\sin\phi)=
		%\begin{align}
		\int_{\mathbb{R}^{+}}\int_{0}^{2\pi}\mathcal{K}_{1}(l\cos\phi,l\sin\phi,r,\theta)f_{R}(r)f_{\theta}(\theta)d\theta dr,$}
	%\end{align} 
	\emph{with}
	\begin{align}
		\mathcal{K}_{1}(l\cos\phi,l\sin\phi,r,\theta)\!=\!
		\exp\Bigg({-\lambda_{t}\!\bigintsss_{\mathbb{R}}1\!-\!\Bigg(\Big(1\!+\!\frac{s (f^2(l\cos\phi,l\sin\phi,u,r,\theta)+\!h^2)^{-\frac{\alpha}{2}} }{m}\Big)^{-m}\Bigg)\!dt}\Bigg). 
	\end{align}
	\begin{proof}[Proof:\nopunct] See Appendix \ref{appendix_A}.
	\end{proof} 
\end{lemma}
By choosing the proper function $f_{R}(r)$ in Lemma \ref{lemma2}, we can calculate the Laplace transform of the interference when the distance between the corridor and its CP follows a truncated Gaussian distribution or a uniform distribution. 
Next, we use the Laplace transform of the interference obtained in Lemma \ref{lemma2} to determine the connectivity performance of UAM air-to-ground communications and the staleness distribution of locally trained FNN parameters.
\subsection{Connectivity Probability and Staleness Distribution}
The connectivity probability is defined as the probability
with which the SIR received by the typical GBS exceeds a target
threshold $\gamma$ required for a successful communication.
Based on the Laplace transform of the interference obtained in Lemma \ref{lemma2},
we can derive the mathematical expression for the connectivity
probability in the following theorem.
\begin{theorem}
	\label{theorem}
	\emph{When an arbitrarily selected aircraft communicates with its associated GBS in UAM, the connectivity probability can be calculated as \vspace{0.05in}}
	\begin{align}
		\mathbb{P}_{\emph{conn}}=\int_{\mathbb{R}^{+}}\sum_{\hat{m}=1}^{m}(-1)^{\hat{m}+1}{m\choose \hat{m}} \mathcal{L}(\gamma\hat{m}\eta(h^2+\beta^2)^{\frac{\alpha}{2}})f_{B}(\beta)d\beta,
	\end{align}
	\emph{where $\eta = m(m!)^{-1/m}$.}
	\begin{proof}[Proof:\nopunct] See Appendix \ref{appendix_B}. 
	\end{proof}
\end{theorem}
From Theorem \ref{theorem}, we can theoretically analyze the connectivity performance of aircraft-to-GBS communication networks in  UAM.
The results in Theorem 1 also pave the way for optimizing the UAM design to achieve a reliable wireless connectivity and ensure an efficient and safe UAM operation. 
%The staleness of the received local FNN parameters is defined as how much time elapsed when the GBS receives the model parameter update from the local aircraft. 
Based on the Theorem \ref{theorem}, we can derive the mathematical distribution of staleness associated to the FNN parameters, as shown in the following corollary.

\begin{corollary}
	\label{corollary1}
	\emph{When an arbitrarily selected UAM aircraft participates in the AFL, the staleness of FNN parameters received at the GBS will follow a cumulative distribution function (CDF) as follows:}
	\begin{align}
		\mathbb{P}_{\emph{staleness}}(\delta \leq \tau )=\int_{\mathbb{R}^{+}}\sum_{\hat{m}=1}^{m}(-1)^{\hat{m}+1}{m\choose \hat{m}} \mathcal{L}\left(\hat{m}\eta\left(2^{\frac{V}{W(\tau-t_{\emph{comp}})}}-1\right)(h^2+\beta^2)^{\frac{\alpha}{2}}\right)f_{B}(\beta)d\beta.
	\end{align}
	\begin{proof}[Proof:\nopunct] We replace $\gamma$ in Theorem \ref{theorem} with $2^{\frac{V}{W(\tau-t_{\text{comp}})}}\!-\!1$ and the remaining proof is similar to Theorem \ref{theorem}. 
	\end{proof}
\end{corollary}

From Corollary \ref{corollary1}, we can observe how different UAM parameters, such as the densities of corridors, aircraft, CPs, and GBSs, and wireless parameters, like the Nakagami fading parameter, affect the staleness distribution of the local FNN model updates received at the GBS. 
To determine the convergence performance of AFL, we still need to derive the number of UAM aircraft that collaboratively learn the FNN-based turbulence prediction model. 

\subsection{Expected Number of Aircraft Participating in AFL}
To determine the expected number of UAM aircraft participating in AFL, we consider a circular area with a radius $R$ and, in the following lemma, we derive the intermediate results, i.e., the expected length of corridor existing in the circular area. 
\begin{lemma}
	\label{circular}
	\emph{For a circular area with radius $R$, the expected length of an arbitrarily selected corridor in UAM is} 
	\begin{align}
		\label{r}
		\mathbb{E}(L) = \int_{\mathbb{R}^{+}} \int_{0}^{2\pi}2 \sqrt{R^2- (x_{1}\cos\theta+x_2\sin\theta+r)^2} f_{R}(r)f_{\theta}(\theta)d\theta dR.
	\end{align}	
	\begin{proof}[Proof:\nopunct]
		See Appendix \ref{Appendix_C}.
	\end{proof}
\end{lemma}
Given the expected length of corridor calculated in Lemma \ref{circular}, we can determine the density of PPP distributed aircraft on each corridor and further obtain the expected number of aircraft associated to the GBS and participating in the AFL in the following theorem. 

\begin{theorem}
	\label{theorem_2}
	\emph{When UAM aircraft operate in a circular area with radius $R$, the expected number $K$ of aircraft participating  in the AFL framework will be given by:}
	\begin{align}
		K = \floor*{\frac{\lambda_{c}\lambda_{t}\mathbb{E}(L)\sum_{n=0}^{N}n\mathbb{P}(n|n<N)}{\lambda_{b}}},
	\end{align}
	\emph{where $\floor*{.}$ is the floor function.}
	\begin{proof}[Proof:\nopunct]
		See Appendix \ref{Appendix_D}.
	\end{proof}
\end{theorem}
From Theorem \ref{theorem_2}, we can determine the number of aircraft participating in AFL to collaboratively learn the turbulence prediction model. 
Next, based on the results in Corollary \ref{corollary1} and Theorem \ref{theorem_2}, we propose a staleness-aware AFL framework for UAM aircraft turbulence prediction and study its convergence.
%propose the staleness-aware AFL and study the convergence based on the results given in Corollary  and Theorem .

\section{Staleness-aware AFL for UAM Aircraft Turbulence Prediction Optimization}
%\subsection{Motivation of Asynchronous FL}
%Although the conventional FL algorithms, like FedAvg \cite{konevcny2016federated} and FedProx \cite{li2018federated}, are gaining popularity among different applications, all these FL algorithms are synchronous. 
%In particular, in the FedAvg, the new communication round will not start unless all local users finish the model training and transmission. 
%Such method will suffer from straggler effect where the aircraft who finish the training and transmission in an early stage have to wait the counterpart that takes a longer time, inevitably increasing the convergence time and jeopardizing the aircraft's ability to quickly predict the turbulence. 
%Different from FedAvg, the FedProx will update the global FNN model by solely considering the model parameters learned by the first certain number of aircraft in each communication round. 
%In this case, the aircraft around the associated GBS are more likely to contribute their learned model parameters to the FL training, and due to the relatively large communication distance, the aircraft on the edge can be totally ignored. 
%However, it is clear that the aircraft on the edge are usually the first affected by the environmental changes, like wind gusts and pressure changes, and thereby, the knowledge learned by these aircraft are of great importance for other aircraft (e.g., aircraft around the GBS) for the turbulence prediction. 
%Therefore, synchronous FL algorithms are not suitable when the UAM aircraft use the framework of FNN and FL for the turbulence prediction.

To mitigate the impact of stale local FNN parameters on the overall convergence, we consider a staleness-aware global aggregation scheme for AFL where the GBS will aggregate the local trained parameters and update the global FNN model as follows 
\begin{align}
	\label{10relationship}
	\boldsymbol{w}_{i+1} = \boldsymbol{w}_{i} - \eta_{i} g_{1}(\delta_{k,i}) \frac{s_{k}}{s_{K}} \nabla f_{k}(\boldsymbol{w}_{\delta_{k,i}}). 
\end{align}
In (\ref{10relationship}), $\eta_{i}$ is the learning rate at communication round $i$, and $\delta_{k,i}$ captures the staleness of FNN parameters received at the GBS from aircraft $k \in \mathcal{K}$ at communication round $i$.
$g_{1}(\delta_{k,i})$ is a monotonically decreasing function of the staleness $\delta_{k,i}$ so as to minimize the impact of stale FNN parameters on the learning performance.
It is clear that the convergence of the turbulence prediction model depends on the expression of the monotonically decreasing function $g_{1}(.)$ and the number of UAM aircraft participating in AFL. 
In the following subsection, we will determine the convergence conditions and analyze the convergence rate of AFL framework with the staleness-aware global aggregation scheme in (\ref{10relationship}).
\subsection{Convergence Study of the Proposed AFL}
Unlike the convergence study done in \cite{xie2019asynchronous} where the staleness of local model updates is bounded, we will explicitly consider the randomly distributed and unbounded staleness derived in Corollary \ref{corollary1} and its impact on the overall AFL convergence. 
To this end, we make the following assumptions: 
\begin{itemize}
	\item The gradient function $\nabla f(.)$ is uniformly Lipschitz continuous, i.e., for some positive parameter $L$, $||\nabla f(\boldsymbol{w}_{i}) - \nabla f(\boldsymbol{w}_{j})|| \leq L ||\boldsymbol{w}_{i} - \boldsymbol{w}_{j}||$. 
	\item The variance of the local gradient descent at an arbitrarily selected aircraft $k \in \mathcal{K}$ with respect to the counterpart for the whole training data across all aircraft is upper bounded, i.e., $\mathbb{E}||\nabla f(\boldsymbol{w})- \nabla f_{k}(\boldsymbol{w})|| \leq \phi^{2}, \forall \boldsymbol{w} \in \mathbb{R}^{d}$, where $\phi^2$ is the upper bound.
	\item The variance of the global gradient descent at communication round $i$ with respect to the local gradient descent of the stale local FNN parameters $\boldsymbol{w}_{\delta_{k,i}}$ at aircraft $k$ is upper bounded by a staleness-dependent value, i.e., $\mathbb{E}[||\nabla f(\boldsymbol{w}_{i})- \nabla f(\boldsymbol{w}_{\delta_{k,i}})||^2] \leq g_{2}(\delta_{k,i}) \mathbb{E}[||\nabla f(\boldsymbol{w}_{\delta_{k,i}})||^2]$, where $g_{2}(\delta_{k,i})$ is a monotonically increasing function in terms of staleness $\delta_{k,i}$ with $g_{2}(0)=0$ and $g_{2}(\infty)=1$.
\end{itemize}

The first two assumptions are commonly used in the current literature, like \cite{doi:10.1137/16M1080173}. In particular, the first assumption can be easily satisfied by some popular loss functions used for the turbulence prediction, such as the mean squared error \cite{10.1145/3394486.3403198,kochkov2021machine,li2020fourier}. 
The second assumption can be justified by the fact that the turbulence variations are bounded in real scenarios. 
For the third assumption, it originates from the basic convergence pattern in which the expected loss decreases as the number of communication rounds between the GBS and UAM aircraft increases. 
Using these three assumptions, we can derive the convergence rate of the proposed AFL to determine the expected loss reduced between two consecutive communication rounds in the following theorem. 

\begin{theorem}
	\label{theorem_3}
	\emph{When the staleness-aware AFL framework is used to improve the turbulence prediction model for UAM aircraft, the convergence rate can be given by:}
	\begin{align}
		\label{13}
		\mathbb{E}(f(\boldsymbol{w}_{i+1})) \leq f(\boldsymbol{w}_{i})- \frac{\eta_{i}\mathbb{E}_{\delta}(g_{1}(\delta))}{2K} ||\nabla f(\boldsymbol{w}_{i})||^2+\frac{L\eta^{2}_{i}\left(\mathbb{V}_{\delta}(g_{1}(\delta))+\mathbb{E}_{\delta}^2(g_{1}(\delta))\right)\phi^2}{2K^2},
	\end{align}
	\emph{as long as the following condition is satisfied:}
	\begin{align}
		\label{convergence_condition}
		L\eta_{i}g_{2}(\delta_{k,i}) +L\eta_{i}g_{1}(\delta_{k,i})  -  K \leq 0, \forall \delta_{k,i} \in \mathbb{R}_{+},
	\end{align}
	\emph{where the mean is $\mathbb{E}_{\delta}(g_{1}(\delta)) = \int_{0}^{\infty}g_{1}(\tau)f_{\delta}(\tau)d\tau$ and the variance is $\mathbb{V}_{\delta}(g_{1}(\delta))=\int_{0}^{\infty}g^2_{1}(\tau)f_{\delta}(\tau)\\d\tau-\mathbb{E}^2_{\delta}(g_{1}(\delta))$ with $f_{\delta}(.)$ derived in Corollary \ref{corollary1}.}
	\begin{proof}[Proof:\nopunct]
		See Appendix \ref{Appendix_E}.
	\end{proof}
\end{theorem}
Using Theorem \ref{theorem_3}, we can determine the convergence rate and convergence conditions when we use the proposed, staleness-aware AFL scheme to optimize the UAM aircraft turbulence prediction model. 
In particular, the convergence rate depends on the statistical properties (i.e., mean and variance) of randomly distributed staleness linked to local FNN parameters.
Also, there are two convergence conditions: there must be at least $K=\ceil*{L\eta_{i}g_{2}(\delta_{k,i}) +L\eta_{i}g_{1}(\delta_{k,i})}$ UAM aircraft associated with the typical GBS; the monotonically decreasing function $g_{1}(\delta)$ must be chosen in a way that $K\mathbb{E}_{\delta}(g_{1}(\delta)) ||\nabla f(\boldsymbol{w}_{i})||^2\geq L\eta_{i}\left(\mathbb{V}_{\delta}(g_{1}(\delta))+\mathbb{E}_{\delta}^2(g_{1}(\delta))\right)\phi^2$. 
In the following corollary, we also calculate the convergence rate when using the conventional aggregation, i.e., $\boldsymbol{w}_{i+1} = \boldsymbol{w}_{i} - \eta_{i} \frac{s_{k}}{s_{K}} \nabla f_{k}(\boldsymbol{w}_{\delta_{k,i}})$, so as to showcase the benefits of the staleness-aware global aggregation scheme in AFL. 
%To better showcase the benefits of using the staleness-aware global aggregation scheme for AFL, we calculate the convergence rate when using the conventional aggregation, i.e., , in the following corollary.  
\begin{corollary}
	\label{corollary21}
	\emph{If the GBS does not consider the staleness associated to the received local FNN parameters  in the global aggregation process, then, the convergence rate for AFL will be} 
	\begin{align}
		\label{131}
		\mathbb{E}(f(\boldsymbol{w}_{i+1})) \leq f(\boldsymbol{w}_{i})- \frac{\eta_{i}}{2K} ||\nabla f(\boldsymbol{w}_{i})||^2+\frac{L\eta^{2}_{i}\phi^2}{2K^2}.
	\end{align}
	\begin{proof}[Proof:\nopunct]
		We can replace $g_{1}(\delta)=1$ in Theorem \ref{theorem_3} to obtain the convergence rate.
	\end{proof}
\end{corollary}
When $(\mathbb{E}_{\delta}(g_{1}(\delta))-1)||\nabla f(\boldsymbol{w}_{i})|| \geq \frac{L\eta_{i}\phi^2}{K}\left(\mathbb{V}_{\delta}(g_{1}(\delta))+\mathbb{E}_{\delta}^2(g_{1}(\delta))\right)$, AFL with the staleness-aware global aggregation can achieve a faster convergence rate than the conventional AFL by comparing the results in Theorem \ref{theorem_3} and Corollary \ref{corollary21}.
%By comparing Theorem \ref{theorem_3} and Corollary \ref{corollary21}, the AFL with the staleness-aware global aggregation can achieve a faster convergence rate than the conventional AFL when , verifying the merits of considering staleness in the AFL design. 
Based on Theorem \ref{theorem_3}, we further determine how fast the model converges to the optimal model in (\ref{optimal}) when using the proposed AFL framework in the following corollary.
%In the following corollary, we also extend Theorem 3 to further determine.
\begin{corollary}
	\label{corollary2}
	\emph{If the loss function is differentiable and strongly convex with positive parameter $c$ and the learning rate is fixed, i.e., $\eta_{i}=\eta$, the results in Theorem \ref{theorem_3} can be simplified to}
	\begin{align}
		\mathbb{E}(f(\boldsymbol{w}_{i+1}))-f(\boldsymbol{w}^{*}) \leq& \left(1- \frac{\eta\mathbb{E}_{\delta}(g_{1}(\delta))c}{K}\right)^{i+1}\left(f(\boldsymbol{w}_{0}-f(\boldsymbol{w}^{*}))\right) \nonumber \\ &+ \frac{L\eta\left(\mathbb{V}_{\delta}(g_{1}(\delta))+\mathbb{E}_{\delta}^2(g_{1}(\delta))\right)\phi^2}{2K\mathbb{E}_{\delta}(g_{1}(\delta))c}\left(1-\left( 1- \frac{\eta\mathbb{E}_{\delta}(g_{1}(\delta))c}{K}  \right)^i\right),
	\end{align}
	\emph{where $\boldsymbol{w}^{*}$ is the solution to the optimization problem in (\ref{optimal}).}
	\begin{proof}[Proof:\nopunct]
		See Appendix \ref{Appendix_F}.
	\end{proof}
\end{corollary}
The result in Corollary \ref{corollary2} shows that, as communication round $i$ increases, there will be a gap, $\frac{L\eta\left(\mathbb{V}_{\delta}(g_{1}(\delta))+\mathbb{E}_{\delta}^2(g_{1}(\delta))\right)\phi^2}{2K\mathbb{E}_{\delta}(g_{1}(\delta))c}$, between $\mathbb{E}(f(\boldsymbol{w}_{i+1}))$ and $f(\boldsymbol{w}^{*})$. 
Clearly, this gap depends on the distribution of staleness associated to the local FNN parameters and the number of UAM aircraft participating in AFL. For instance, a larger $K$ will lead to a smaller gap between $\mathbb{E}(f(\boldsymbol{w}_{i+1}))$ and $f(\boldsymbol{w}^{*})$. 
Hence, to minimize the gap derived in Corollary \ref{corollary2} and improve the convergence performance, we can optimize the UAM parameter setting (e.g., select a proper density of corridors) and the wireless network design (e.g., transmit power).

\begin{table}[!t]
	\normalsize
	%\large
	\begin{center}
		%\centering
		\caption{ Simulation parameters.}
		\label{table_example}
		\resizebox{12cm}{!}{
			\begin{tabular}{|c|c|}
				\hline
				\textbf{Parameters} & \textbf{Values} \\ \hline	
				Height $h$ & $500$~feet ($152.4$~m) \cite{FAA} \\ \hline
				CP density $\lambda_{c}$     & $0.001$~(km$^{2}$)$^{-1}$    \\ \hline
				GBS density $\lambda_{b}$ & $1$~(km$^{2}$)$^{-1}$ \\ \hline 
				Corridor density  $\lambda_{l}$ &  $5$~corridor/km$^{2}$ \\ \hline 
				Aircraft density $\lambda_{t}$ & $1$~aircraft/km \\ \hline 
				Maximum number of corridors $N$ & $10$ \\ \hline 
				Transmit power $T_{r}$ & $40$~dBm \\ \hline 
				SIR threshold $\gamma$  & $0$~dB \\ \hline
				Bandwidth $W$  & $10$~Mhz \\ \hline
				Model parameter size $V$  & $5$~kb \\ \hline
				Path loss exponent $\alpha$  & $4$ \\ \hline
				Nakagami fading parameter $m$  & $1$ \\ \hline
				%Probability that a random aircraft is a transmitter $\rho$ & $0.5$ \\ \hline 
				Variance of truncated Gaussian distribution $\sigma^2$ & $1$ \\ \hline 
				Distance limitation for the uniform distribution $\hat{r}$ & $2$~km \\ \hline  
				Training data size $v$ & $10^3$~bits  \\ \hline 
				Number of computing cycles needed per bit $\epsilon$ & $10^3$  \\ \hline  Frequency of CPU clock $\varepsilon$ &  $10^9$~cycles/s\\ \hline
		\end{tabular}}
	\end{center}
%\vspace{-0.1in}
\end{table}

\section{Simulation results}
For our simulations, we model the UAM system as a circular area with a radius of $20$~km. Simulation parameters are summarized in Table \ref{table_example}.
We first validate the theoretical analysis in Theorem \ref{theorem} for the two discussed cases where the distance $r$ follows a truncated Gaussian distribution and a uniform distribution. 
Then, we study how the densities of GBSs, CPs, corridors, and aircraft affect the wireless connectivity performance and provide insights into the system design guidelines for UAM. 
Next, we analyze the convergence of staleness-aware AFL framework in comparison to multiple baselines and highlight the merits of considering staleness in the AFL framework. 

\begin{figure}[!t]
	\centering 
	\subfloat[Truncated Gaussian distribution for the distance \newline $r$ between the CP and corridors.]{%
		\includegraphics[width=2.8in,height=1.8in]{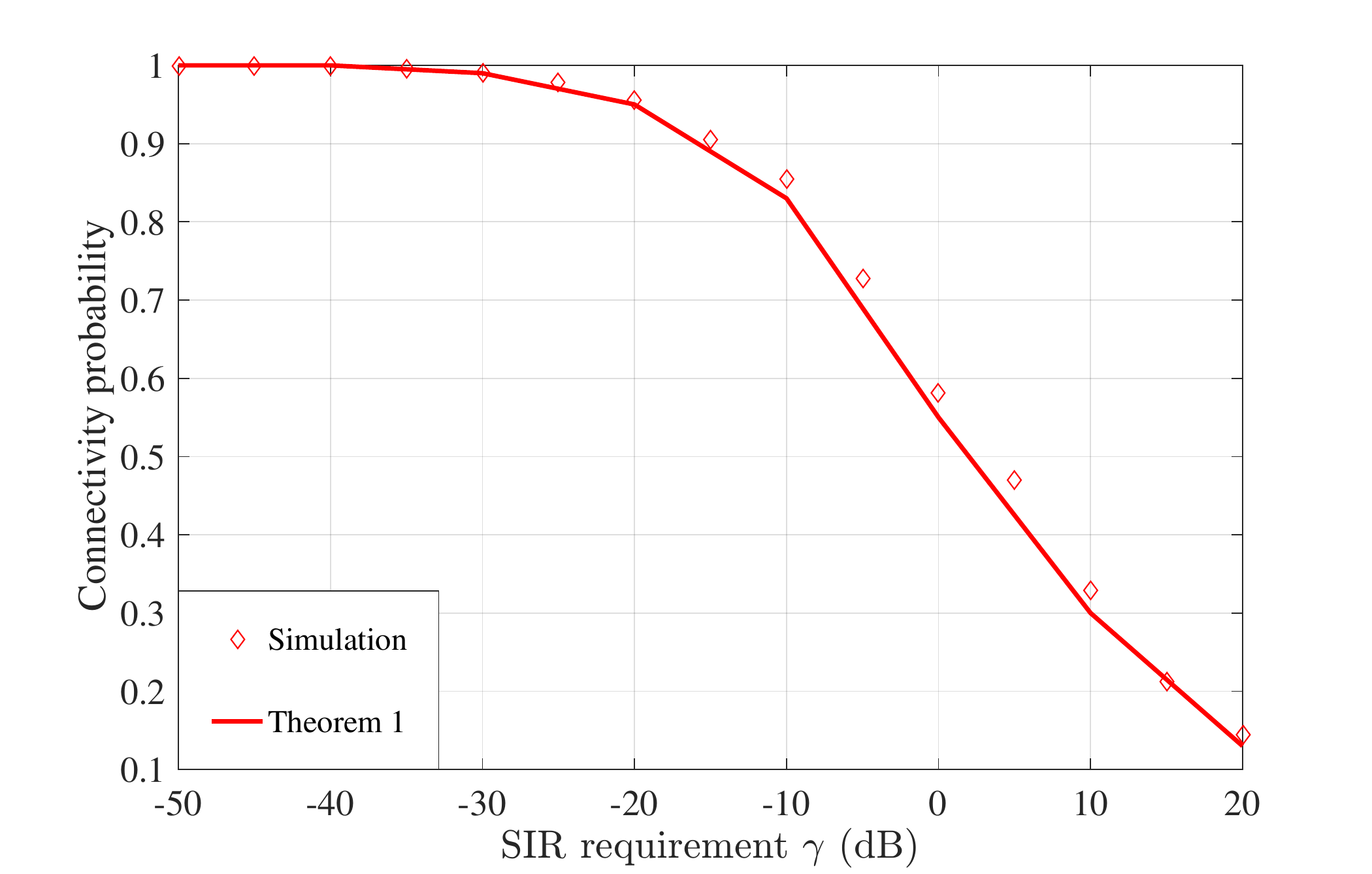}\label{21}
	}	
	\subfloat[Uniform distribution for the distance $r$ between the CP and corridors.]{%
		\includegraphics[width=2.8in,height=1.8in]{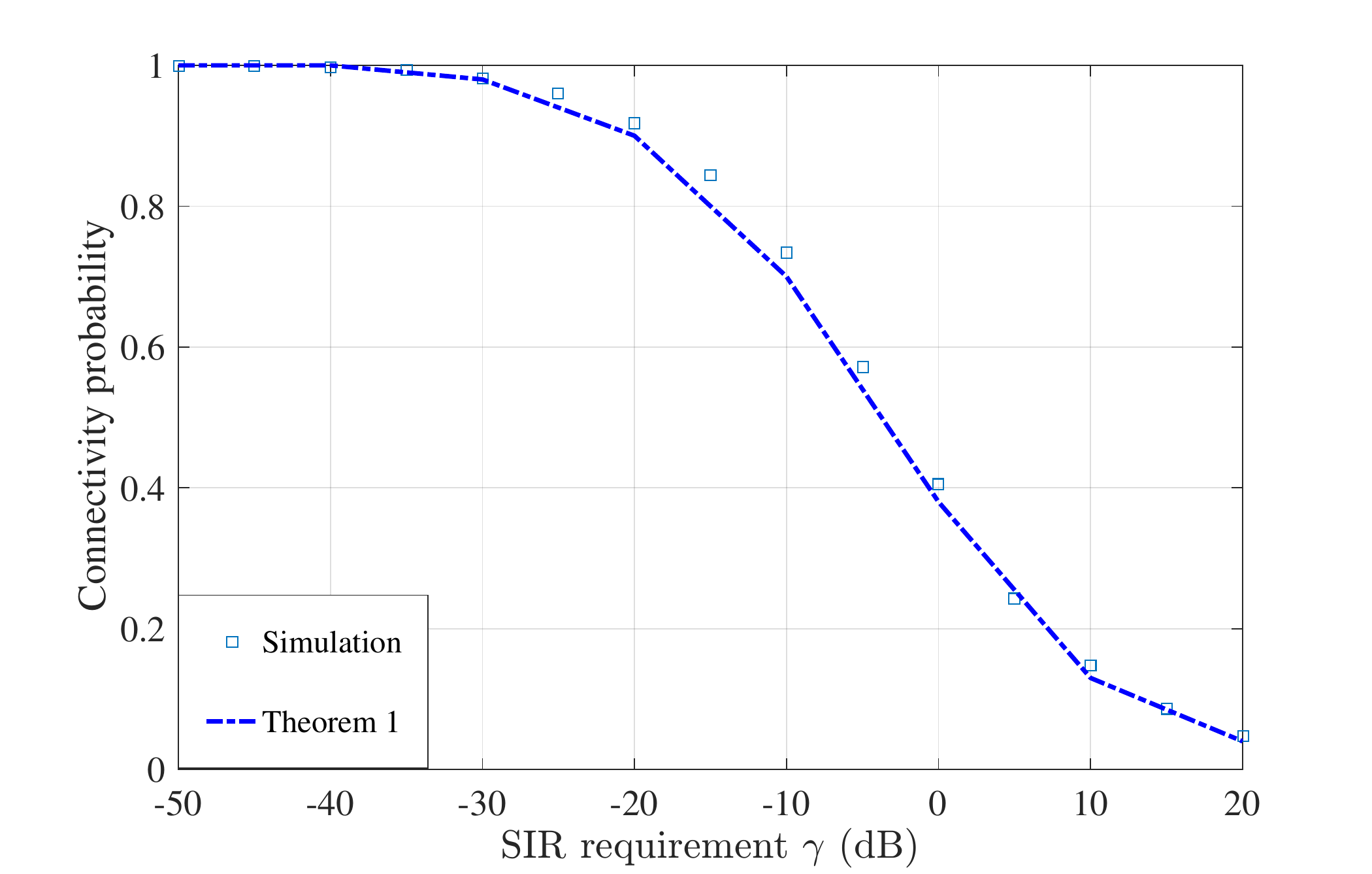}\label{22}
	}	
	\caption{Connectivity probability of the UAM wireless network versus the SIR threshold.}
	\label{Fig2}	
	%\vspace{-0.1in}
\end{figure}

\subsection{Validation of Theoretical Analysis}
Fig. \ref{Fig2} shows the connectivity probability of the air-to-ground communication network versus the SIR threshold $\gamma$ when the distance $r$ follows a truncated Gaussian distribution and a uniform distribution. 
As observed from Fig. \ref{Fig2}, the simulation results for both distributions match the analytical results with a small deviation.
The small deviation stems from the use of the approximated tail probability of the Gamma function in Theorem \ref{theorem}.
Fig. \ref{Fig2} also shows that, when the SIR threshold $\gamma$ increases, the connectivity performance of the UAM's wireless network will degrade. 
This is because, with a higher target SIR threshold, fewer air-to-ground communication links will meet the connectivity requirement for a successful transmission. 
%Next, we show the connectivity performance of UAM under different parameter settings.
%that the proposed AFL framework can achieve a fast convergence than the synchronous FL baselines.
%Finally, we highlight the importance of considering the staleness in the AFL framework design and show the convergence performance of the AFL framework under different UAM parameters. 
\subsection{Connectivity Performance of UAM under Different Parameter Settings}

\begin{figure}
	\centering	
	\begin{minipage}{0.43\textwidth}
		\centering
		\includegraphics[width=1\linewidth]{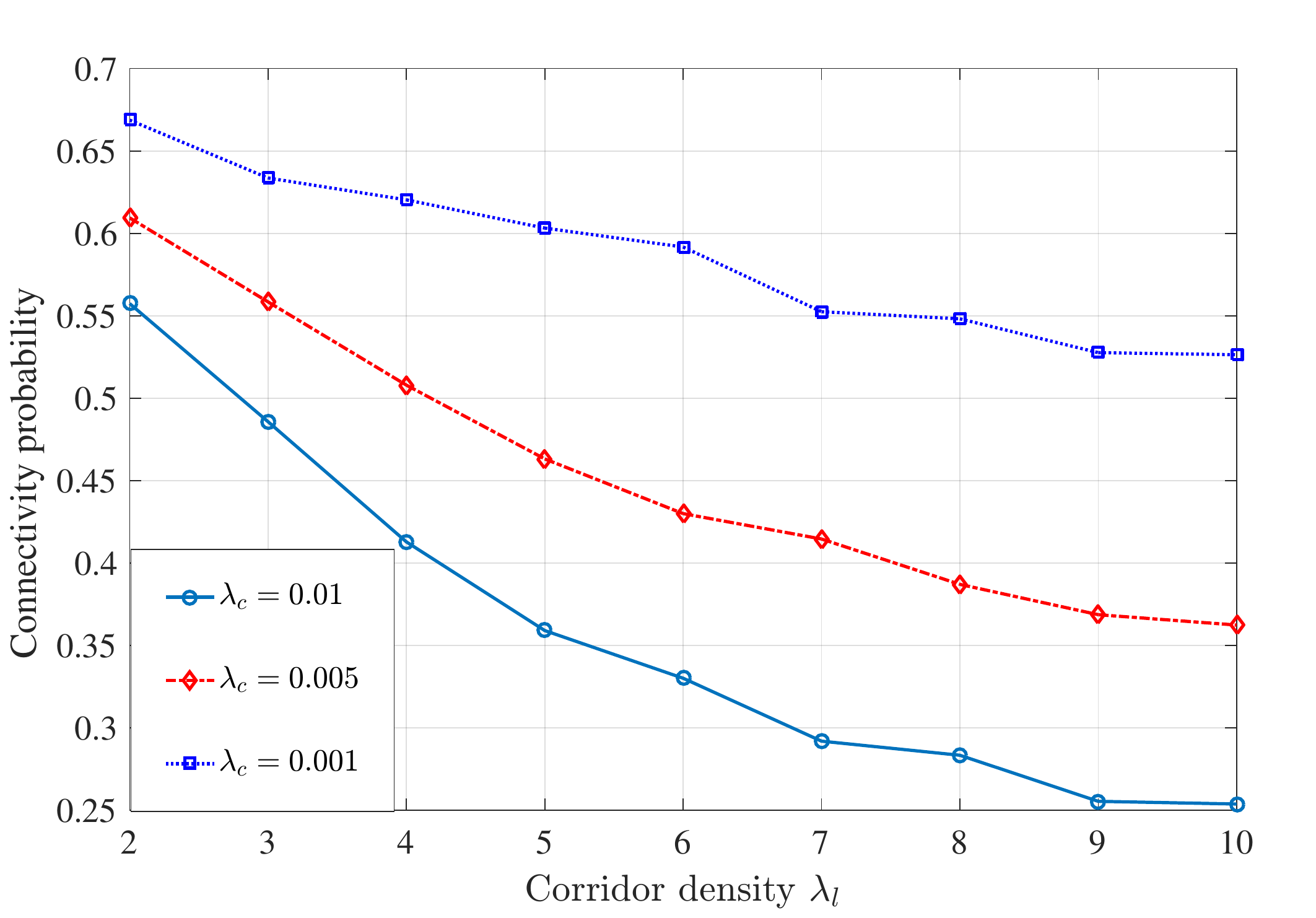}
		\caption{Connectivity probability versus the corridor density under different cluster densities.}
		\label{Fig3}
	\end{minipage}
	\begin{minipage}{0.43\textwidth}
		\centering
		\includegraphics[width=1\linewidth]{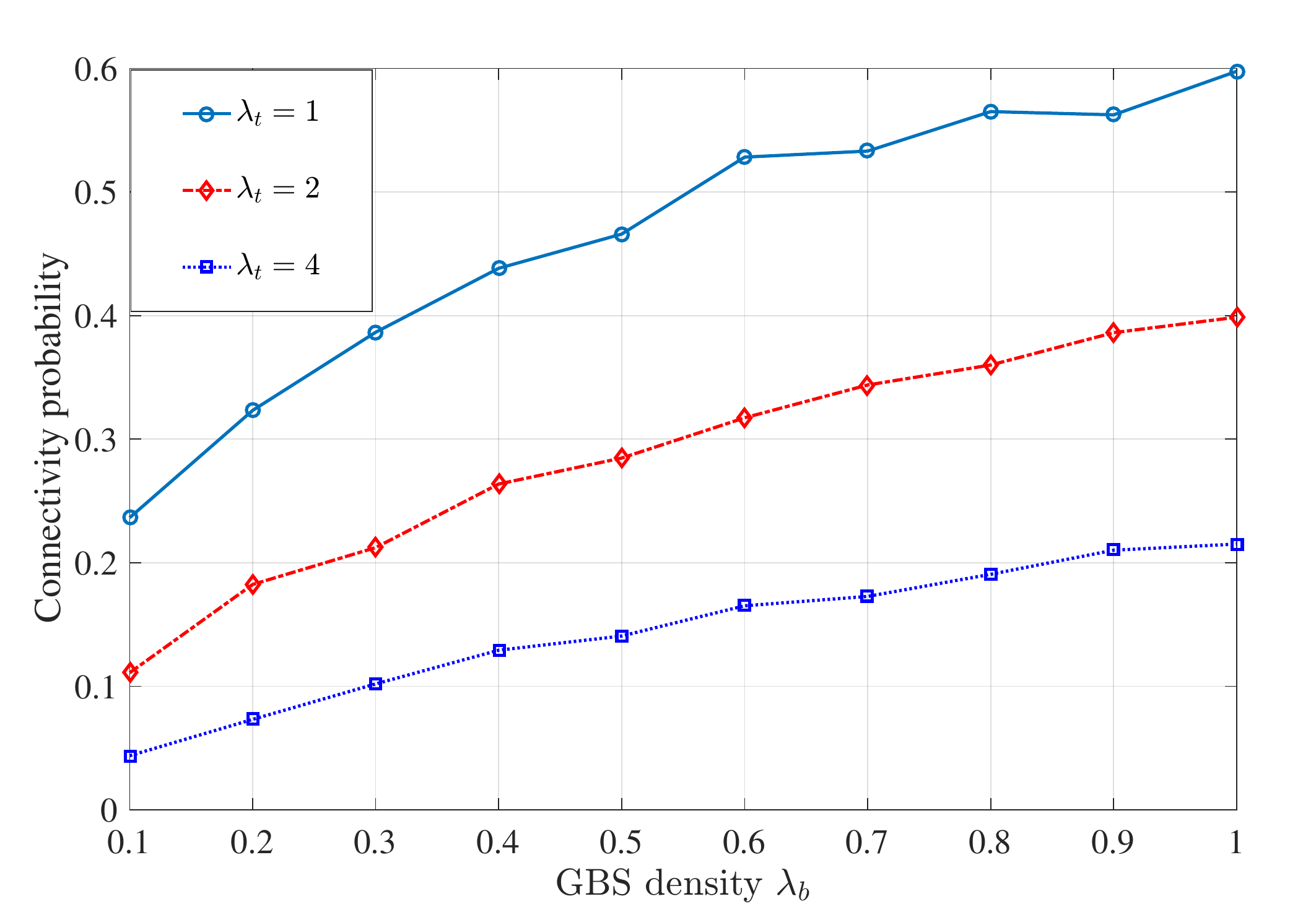}
		\caption{Connectivity probability versus the GBS density under different aircraft densities.}
		\label{Fig4}
	\end{minipage}%%%
%\vspace{-0.1in}
\end{figure}

Fig. \ref{Fig3} shows the connectivity probability of the UAM wireless network versus the corridor density $\lambda_{l}$ under different values for the cluster density $\lambda_{c}$ when the distance follows a truncated Gaussian distribution. 
From Fig. \ref{Fig3}, we observe that the connectivity probability decreases when $\lambda_{l}$ increases. 
This is due to the fact that, with more corridors distributed around the CPs, the number of interfering aircraft will also increase, leading to a higher interference at the typical GBS and a degradation of the wireless connectivity performance.  
Moreover, Fig. \ref{Fig3} shows that a higher density of CPs will also degrade the UAM wireless connectivity performance. 
This is because a higher density of CPs will lead to more corridors and more interfering aircraft.

%\begin{figure}[!t]
%	\centering
%	\includegraphics[width=2.7in,height=2in]{Fig3.pdf}
%	\DeclareGraphicsExtensions.
%	
%	\caption{C}
%	\label{Fig3}\vspace{-0.3in}
%\end{figure} 
%
%\begin{figure}[!t]
%	\centering
%	\includegraphics[width=2.7in,height=2in]{Fig4.pdf}
%	\DeclareGraphicsExtensions.
%	
%	\caption{Connectivity probability
%		of the UAM wireless network versus the GBS density under different aircraft densities.}
%	\label{Fig4}vspace{-0.3in}
%\end{figure} 

Fig. \ref{Fig4} shows the connectivity probability of the UAM wireless network versus the GBS density $\lambda_{b}$ under different aircraft density $\lambda_{t}$ on each corridor and for the case when the distance follows a truncated Gaussian distribution.  
As shown in Fig. \ref{Fig4}, the connectivity probability increases when the GBS density $\lambda_{b}$ increases. 
This is because, with more GBSs, the communication distance between a given GBS and its associated UAM aircraft will be reduced, leading to a higher received signal power and SIR.   
Moreover, Fig. \ref{Fig4} shows that the presence of more aircraft on corridors will negatively impact the connectivity probability performance. 
This can be explained that a higher $\lambda_{t}$ will lead to a larger number of interfering aircraft. 
Note that the simulation results of the uniformly distributed distance $r$ are similar to the counterparts when the distance follows a truncated Gaussian distribution, and they are omitted due to space limitations.

Based on the simulation results in Figs. \ref{Fig3} and \ref{Fig4}, when the densities of aircraft, corridors, and CPs increase, the UAM wireless connectivity performance will degrade. 
To avoid this performance degradation, one can deploy more GBSs at the expense of  site acquisition constraints and costs, particularly in urban environments.
Therefore, realizing efficient UAM systems depends on a careful selection of design parameters to improve the wireless connectivity performance while reducing the overall deployment cost.

\subsection{Convergence of AFL for Turbulence Prediction}
In our simulation, when building the AFL framework, we generate the local training turbulence data for UAM aircraft by first randomly generating the velocity field $\boldsymbol{v}_{0}(s)$ and, then, following the Burgers' equation \cite{Boritchev_2014} and \cite{miller2015toward}: 
\begin{align}
	\label{weight}
	& \frac{\partial \boldsymbol{v}(s,t)}{\partial t} + \frac{1}{2}\frac{\partial \boldsymbol{v}^2(s,t)}{\partial s} =  \rho \frac{\partial \boldsymbol{v}^2(s,t)}{\partial s \partial s}, \boldsymbol{v}(s,0) = \boldsymbol{v}_{0}(s),   
\end{align}
where $\boldsymbol{v}(s,t)$ is the velocity of the turbulence flow at the location $s$ and time $t$, and $\rho$ is the kinematic viscosity coefficient.
Note that, here, we use Burgers' equation as an example to generate the local training data, our proposed AFL framework can be used to deal with the training data generated by other partial differential equations (PDEs), like Navier-Stokes equation.
%\begin{align}
%	\label{weight}
%	& \frac{\partial \boldsymbol{v}(s,t)}{\partial t} + \frac{1}{2}\frac{\partial \boldsymbol{v}^2(s,t)}{\partial s} =  \rho \frac{\partial \boldsymbol{v}^2(s,t)}{\partial s \partial s},\\
%	\label{force}
%	& \nabla \boldsymbol{v}(s,0) = \boldsymbol{v}_{0}(s),   
%\end{align}
%where $\boldsymbol{v}(s,t)$ is the velocity of the air flow at the location $s$ and  time $t$, and $\rho$ is the kinematic viscosity coefficient.
In terms of the turbulence prediction model, we consider an FNN model with two Fourier layers. 
We randomly select a GBS and its associated aircraft to collaboratively learn the FNN model within the proposed AFL framework.  

\begin{figure}[!t]
	\centering 
	\subfloat[First synchronous FL baseline.]{%
		\includegraphics[width=2.1in,height=1.5in]{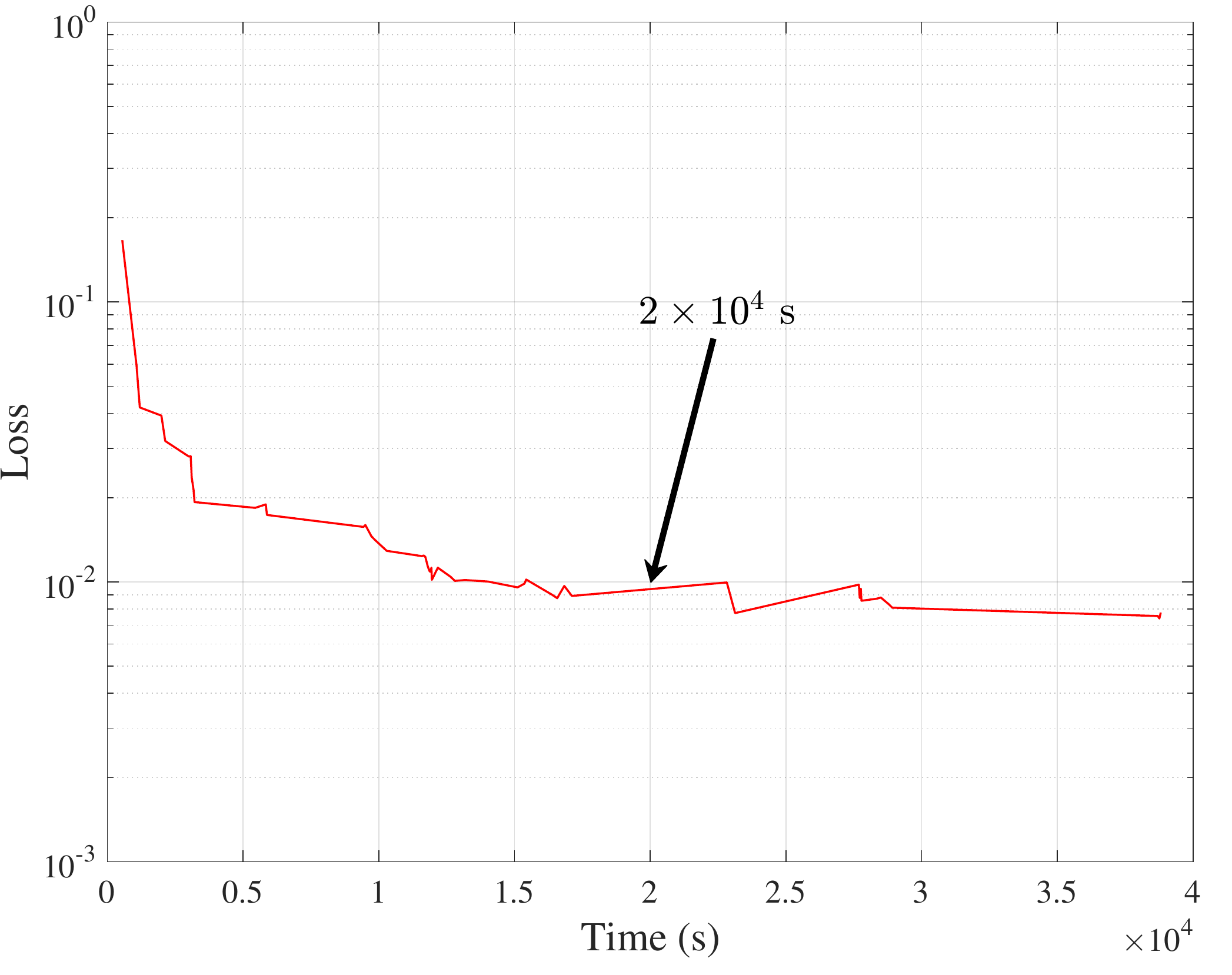}\label{81}
	}	
	\subfloat[Second synchronous FL baseline.]{%
		\includegraphics[width=2.1in,height=1.5in]{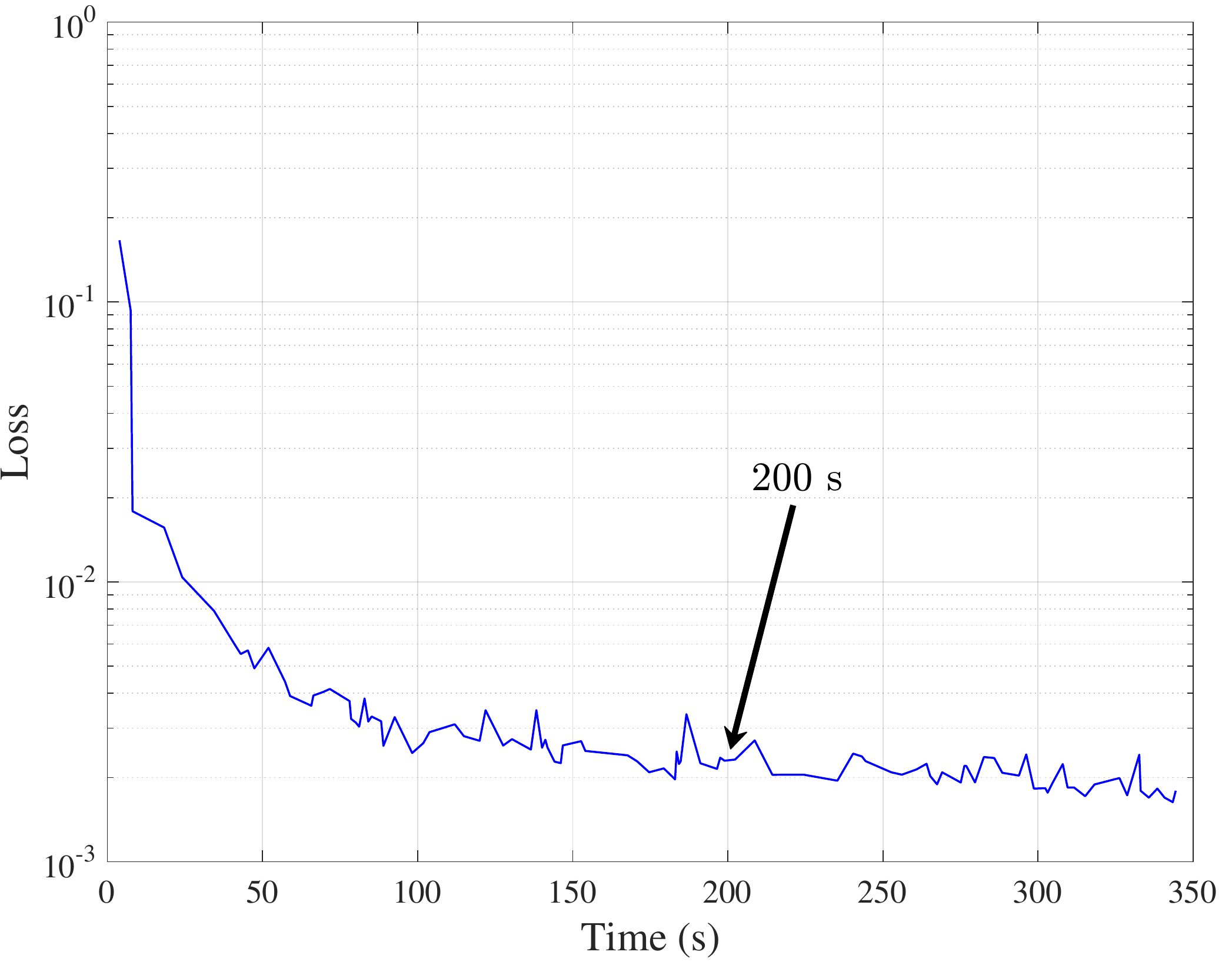}\label{82}
	}
	\subfloat[Proposed AFL framework.]{%
		\includegraphics[width=2.1in,height=1.5in]{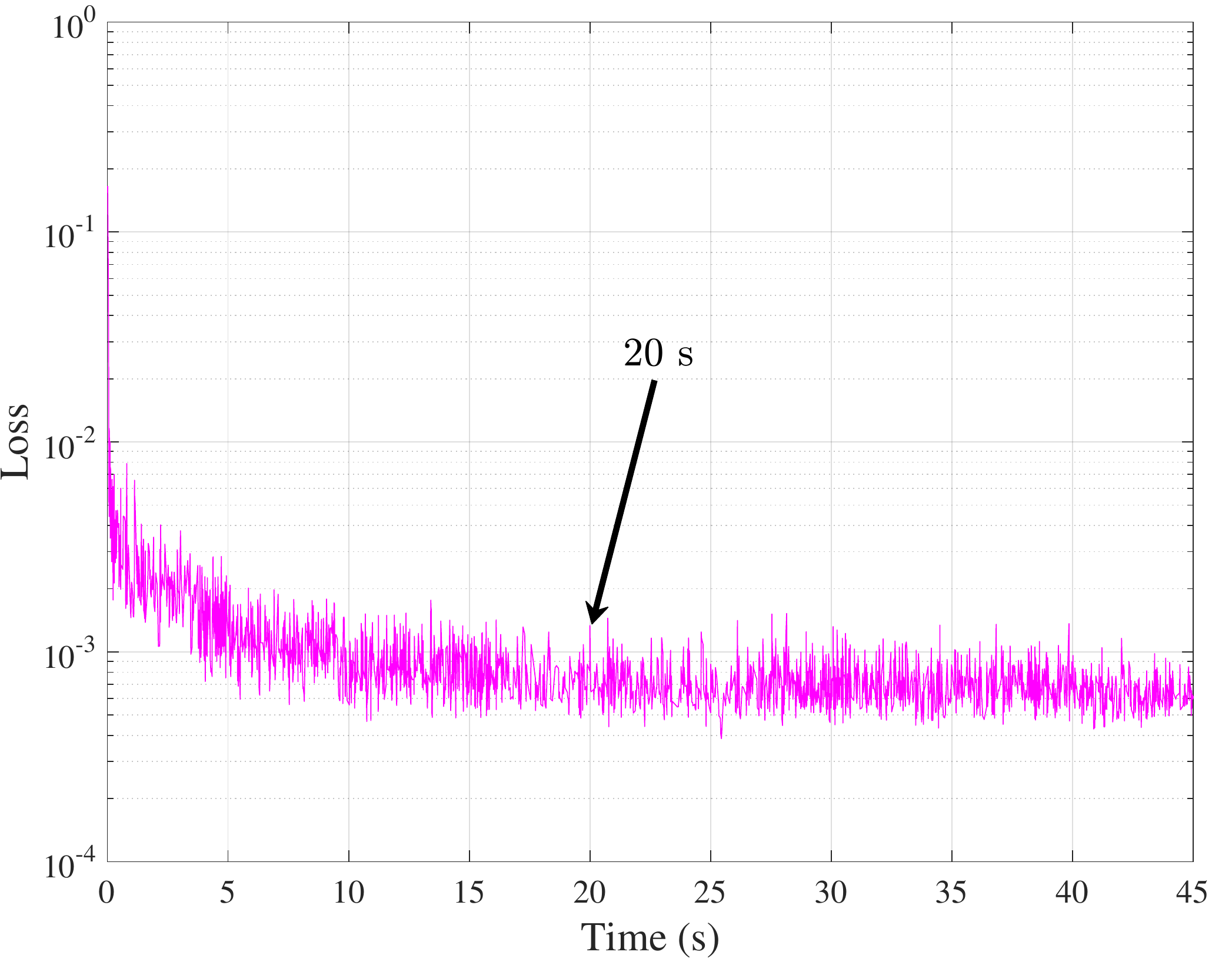}\label{83}
	}	
	\caption{ Training performance of two synchronous FL baselines and our proposed AFL framework.}
	\label{Fig8}	
	%\vspace{-0.1in}
\end{figure}

Fig. \ref{Fig8} shows the convergence of synchronous FL baselines and our proposed AFL framework. 
In particular, we select two popular synchronous FL baselines. 
In the first baseline, i.e., FedAvg \cite{konevcny2016federated}, the GBS will generate a new global FNN model when all local aircraft finish the local model training and the model parameter transmission. 
In the second baseline, i.e., scalable FL \cite{bonawitz2019towards}, a new global FNN model will be generated at the GBS once it receives the locally trained parameters from half of its associated aircraft. 
As shown from Fig. \ref{Fig8}, our proposed AFL framework can achieve a faster convergence compared to two synchronous counterparts.
In particular, our proposed AFL framework can converge within $20$~s, whereas the first and second baselines need to spend, $2 \times 10^4$~s and $200$~s, to achieve convergence. 
In other words, our proposed AFL framework can improve the convergence rate by $99.9\%$ and $90\%$ compared to these two baselines. 
The reason is that, compared with the first baseline, our proposed framework will not suffer from the straggler effect where the aircraft who finish the local training and transmission fast have to wait for the slower ones. 
Also, our proposed AFL framework guarantees that the local model update from any aircraft will be considered in the global aggregation in contrast to the second baseline where some important local model updates can be discarded due to the long communication time. 
%From Fig. \ref{Fig8}, we can observe that the convergence of the first baseline is smoother than the second baseline and our proposed framework. 
%The reason is that, the first baseline can reduce the variance in the loss function by considering the local updates from all aircraft in each training round, whereas the second baseline and our framework aggregate the local FNN parameters  from, respectively, half of all aircraft and a single aircraft. 

\begin{figure}[!t]
	\centering
	\includegraphics[width=2.8in,height=2.0in]{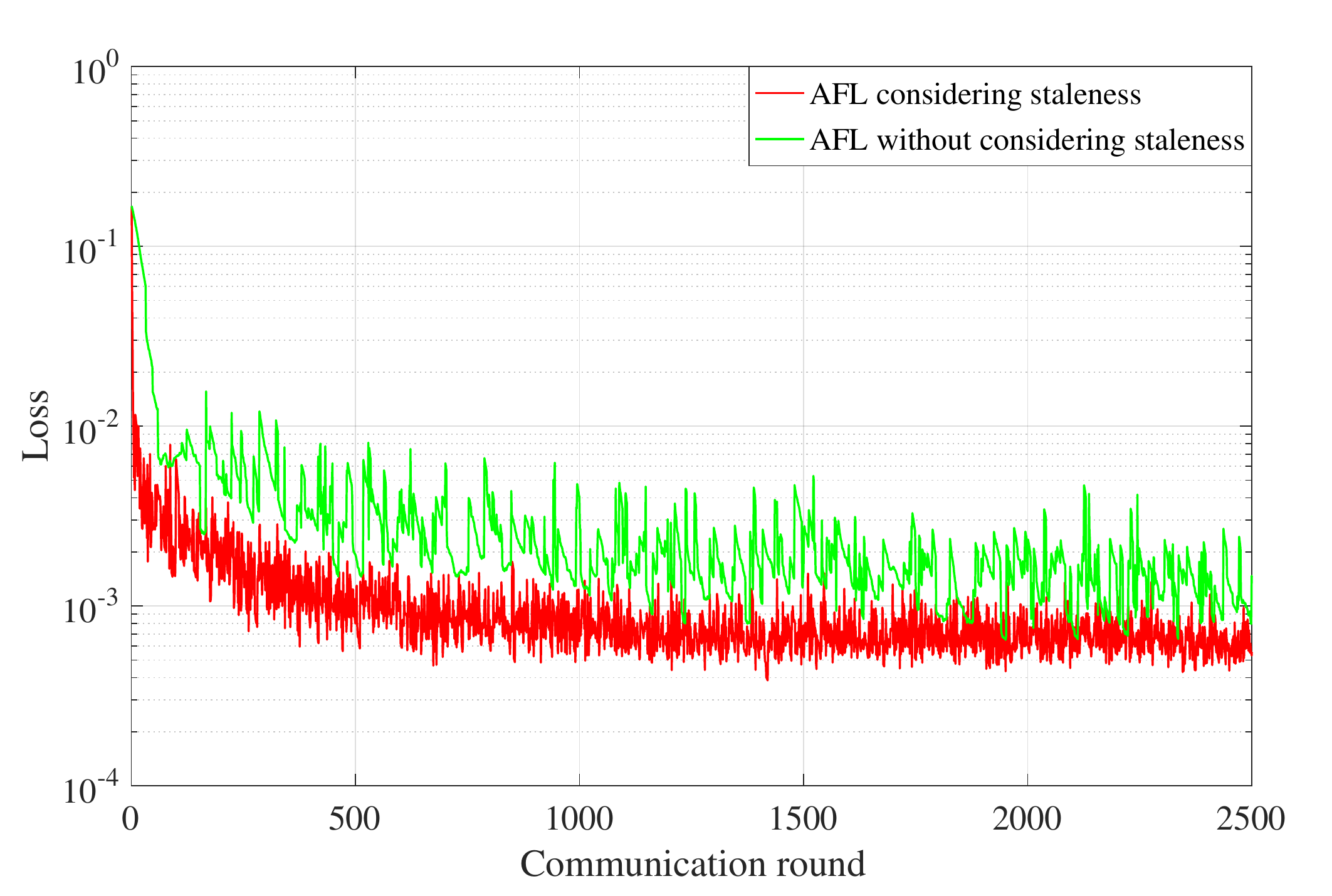}
	\DeclareGraphicsExtensions.
	\caption{Comparison between two AFL schemes with and without considering staleness.}
	\label{Fig9}
	%\vspace{-0.1in}
\end{figure}

In Fig. \ref{Fig9}, we compare our proposed staleness-aware 
AFL framework with an AFL framework that aggregates the local model updates without considering staleness. 
In particular, for the AFL framework, we choose $g_{1}(\delta)= 1+\exp{(-\delta)}$ in (\ref{10relationship}) based on the results derived in Corollary \ref{corollary21}.
In the AFL framework without considering the staleness, $g_{1}(\delta)=1$.
As observed from Fig. \ref{Fig9},  the AFL framework considering the impact of staleness converges faster than the framework without considering the staleness of local model updates. 
In particular, to achieve a loss of $10^{-3}$, our proposed AFL framework needs around $500$ communication rounds, whereas the AFL framework without considering the staleness of local updates needs more than $2000$ communication rounds. 
The reason is that, during the global aggregation in (\ref{10relationship}), the local model update can be generated by using the stale version of global model instead of the freshly generated one at UAM aircraft. 
Hence, directly aggregating such outdated local FNN parameters will inevitably slow down the convergence of the FNN-based turbulence prediction model. 
However, in our proposed AFL framework, we introduce the monotonically decreasing function $g_{1}(.)$ to mitigate the impact of the outdated local parameters on the global aggregation and facilitate the overall  convergence.

\begin{figure}[!t]
	\centering 
	\subfloat[GBS density $\lambda_{b}=0.1$,  corridor \newline density $\lambda_{l}=10$.]{%
		\includegraphics[width=2.1in,height=1.6in]{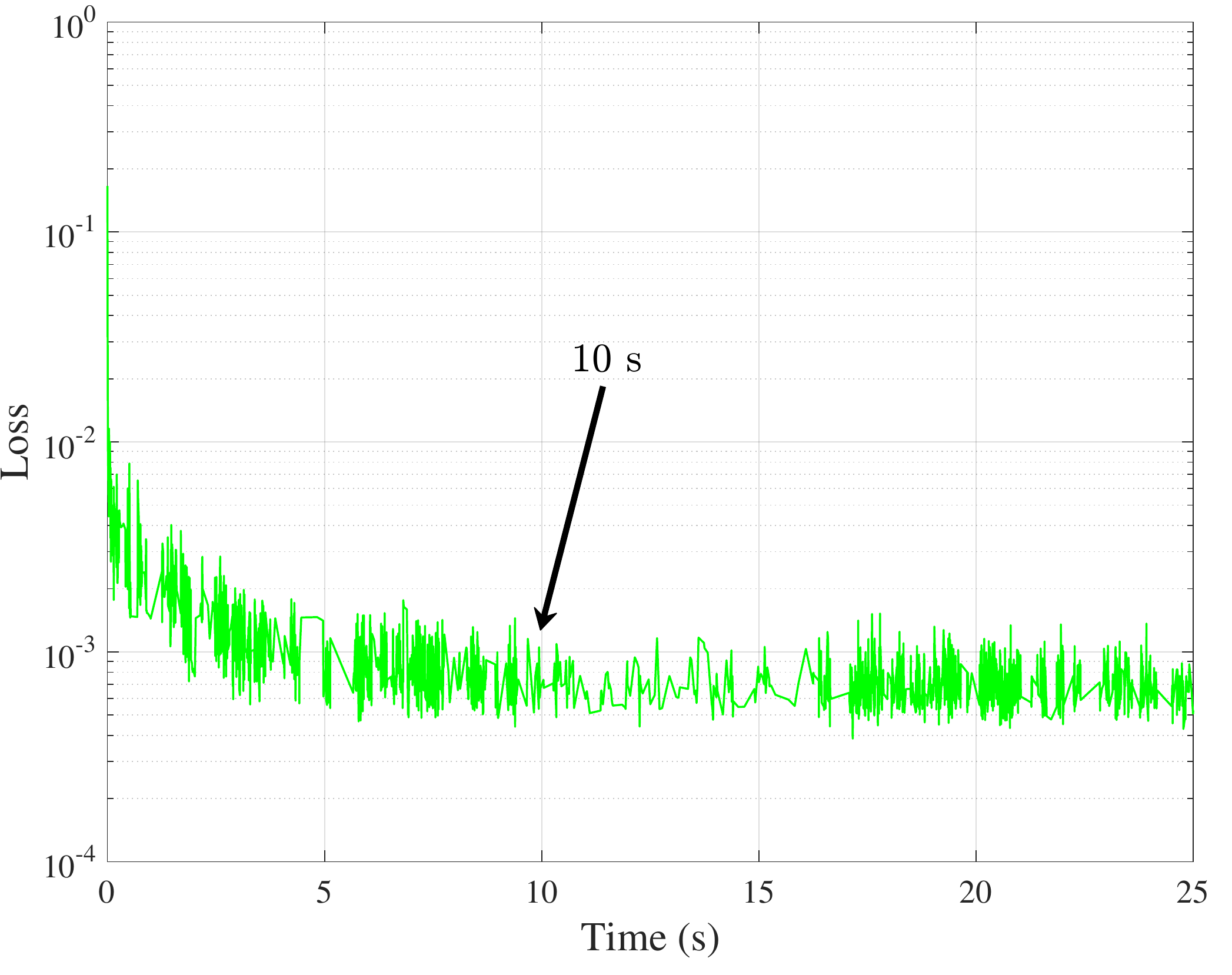}\label{10-1}
	}	
	\subfloat[GBS density $\lambda_{b}=0.05$,  corridor \newline density $\lambda_{l}=10$.]{%
		\includegraphics[width=2.1in,height=1.6in]{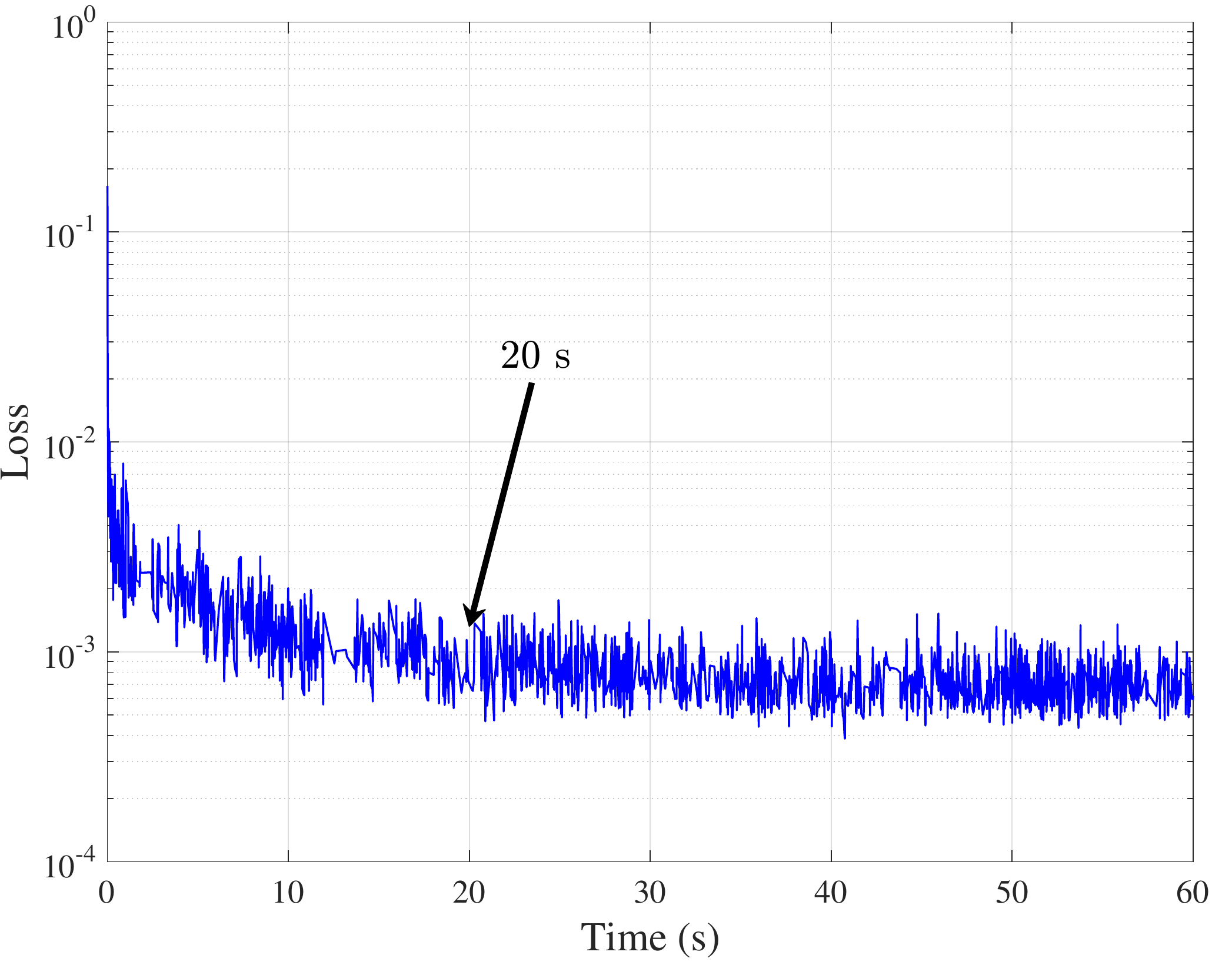}\label{10-2}
	}
	\subfloat[GBS density $\lambda_{b}=0.05$, corridor \newline density $\lambda_{l}=15$.]{%
		\includegraphics[width=2.1in,height=1.6in]{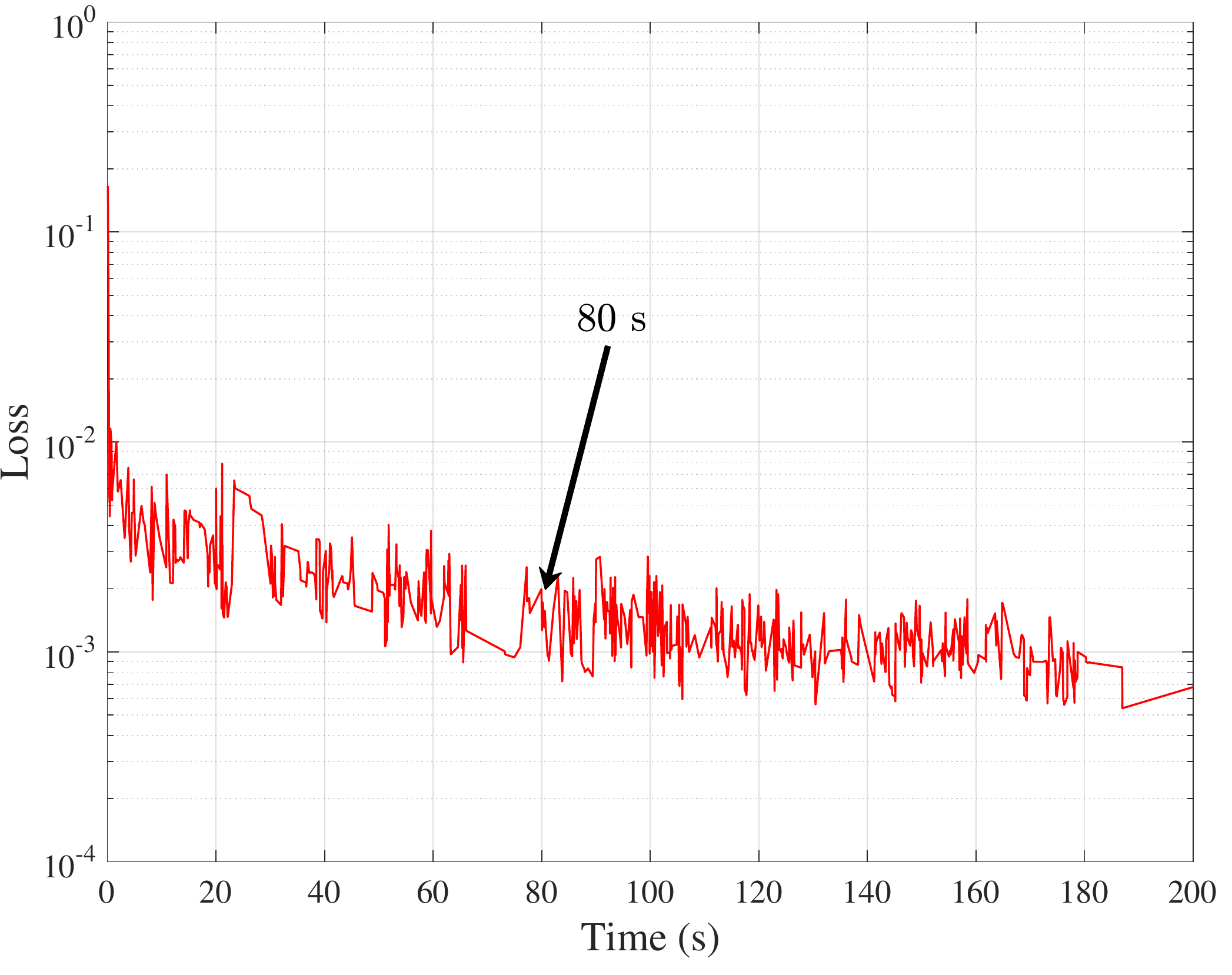}\label{10-3}
	}	
	\caption{Convergence of the proposed AFL under different UAM system settings. These results show that, a high density of GBSs will speed up the convergence, while dense UAM corridors lead to a low convergence rate.}
	\label{Fig10}
	%\vspace{-0.1in}
\end{figure}

%\begin{figure}[!t]
%	\centering
%	\includegraphics[width=3.1in,height=2.2in]{Fig10_update.pdf}
%	\DeclareGraphicsExtensions.
%	\caption{The convergence performance of the proposed staleness-aware AFL under different UAM system settings.}
%	\label{Fig10}vspace{-0.3in}
%\end{figure}
Fig. \ref{Fig10} shows the convergence of the proposed AFL framework under different UAM parameter settings. 
As shown in Fig. \ref{Fig10} \subref{10-1} and \subref{10-2}, when the density of GBS $\lambda_{b}$ increases, the convergence rate of the AFL framework also speeds up (i.e., the convergence time switches from $20$~s to $10$~s). 
The reason is that, a large density $\lambda_{b}$ of GBSs can lead to a better SIR at the receiving GBSs and more frequent uplink-downlink learning model transmission between GBS and UAM aircraft. 
Hence, for a given time period, the total number of communication rounds will increase and the convergence rate will increase. 
Also, with a better SIR, the trained FNN parameters can be quickly transmitted to the GBSs, reducing staleness and improving the convergence. 
Moreover, from Fig. \ref{Fig10} \subref{10-2} and \subref{10-3}, we can observe that as the density of corridors $\lambda_{l}$ increases, the convergence rate becomes slower (i.e., the convergence time switches from $20$~s to $80$~s).  
This is due to the fact that more corridors will lead to more interfering aircraft in the wireless network and the air-to-ground communication link will experience a low SIR. 
Hence, the frequency of FNN parameters transmission between the GBS and UAM aircraft will be low and the staleness of local FNN parameters will also increase, reducing the convergence rate. 
\section{Conclusions}
In this paper, we have characterized the connectivity performance of 
%....use stochastic geometry to model the spatial distribution of GBSs, corridors, and aircraft in UAM, and perform wireless  study for 
the aircraft-to-GBS communication network in UAM.
In particular, we have used PCP and PLP to capture the relative location of corridors around their CPs where their relative distance between them is modeled by two distributions: truncated Gaussian distribution and uniform distribution. 
Next, we have characterized the distribution of aircraft on each corridor and GBSs by using,  respectively, one-dimensional and two-dimensional PPP.  
Using this system setup, we have derived new theoretical results for the SIR-based connectivity probability when the aircraft communicates to its closest GBS and the staleness distribution associated to local FNN parameters.
Based on the wireless connectivity study, we have proposed 
a wireless-enabled AFL framework to collaboratively learn the FNN-based turbulence prediction model among UAM aircraft. 
In particular, a staleness-aware AFL has been introduced to mitigate the impact of staleness associated to local FNN parameters on the overall convergence. 
We have also performed a rigorous convergence study to determine the convergence rate and showcase the merits of considering staleness in the AFL framework design. 
Simulation results corroborate our connectivity analysis for UAM and show how different system settings affect the overall connectivity performance.
Moreover, the results show that the proposed AFL framework outperforms the conventional synchronous FL counterparts and AFL framework without considering the staleness. 
In addition, the results highlight the necessity of optimizing the UAM parameter setting and wireless network design to improve the turbulence prediction model convergence and UAM aircraft wireless connectivity. 
%Future works will extend the proposed staleness-aware AFL framework to the turbulence flow modeled by other PDEs, like Navier-Stokes equation. 
\appendix
\subsection{Proof of Lemma \ref{lemma2}}
\label{appendix_A}
The Laplace transform of interference at the typical GBS can be derived as follows
\begin{align}
	\label{lemma_results}
	&\mathcal{L}(s)\! =\! \mathbb{E}\left[\exp(-s\sum_{i\in\Phi} \sum_{j\in \Psi_{i}}\! \sum_{k \in \Omega_{i,j}}\!\!\! g' (f^2(x_{1},x_{2},u,r,\theta)+h^2)^{-\frac{\alpha}{2}})\right] \nonumber \\ 
	&=\!\mathbb{E}\!\left[\prod_{i\in\Phi}\prod_{j\in \Psi_{i}}\!\prod_{k \in \Omega_{i,j}}\!\!\!\mathbb{E}_{g'}\exp( -sg' (f^2(x_{1},x_{2},u,r,\theta)\!+\!h^2)^{-\frac{\alpha}{2}})\right] \nonumber \\ 
	&\stackrel{(a)}{=}\!\mathbb{E}\!\left[\prod_{i\in\Phi}\prod_{j\in \Psi_{i}}\prod_{k \in \Omega_{i,j}}\left(1\!+\!\frac{s (f^2(x_{1},x_{2},u,r,\theta)\!+\!h^2)^{-\frac{\alpha}{2}} }{m}\right)^{-m}\right] \nonumber \\ 				&\stackrel{}{=}\!\mathbb{E}\!\left[\prod_{i\in\Phi}\prod_{j\in \Psi_{i}}\!\!\mathbb{E}_{\Omega_{i,j}}\!\!\!\prod_{k \in \Omega_{i,j}}\!\!\!\!\left(\!1\!+\!\frac{s (f^2(x_{1},x_{2},u,r,\theta)\!+\!h^2)^{-\frac{\alpha}{2}} }{m}\!\right)^{-m}\right] \nonumber \\ 
	&\stackrel{(b)}{=}\mathbb{E}\left[\prod_{i\in\Phi}\prod_{j\in \Psi_{i}}e^{-\lambda_{t}\int_{\mathbb{R}}1-\left(\left(1+\frac{s (f^2(x_{1},x_{2},u,r,\theta)+h^2)^{-\frac{\alpha}{2}} }{m}\!\right)^{-m}\right)\!dt}\right] \nonumber \\ 
	&\stackrel{}{=}\!\mathbb{E}\!\left[\!\prod_{i\in\Phi}\!\mathbb{E}_{\Psi_{i}}\!\!\prod_{j\in \Psi_{i}}\!\!\underbrace{e^{-\lambda_{t}\int_{\mathbb{R}}1-\left(\left(1\!+\!\frac{s (f^2(x_{1},x_{2},u,r,\theta)+h^2)^{-\frac{\alpha}{2}} }{m}\!\right)^{-m}\right)\!dt}}_{\mathcal{K}_{1}(x_1,x_2,r,\theta)}\right] \nonumber \\ 
	&\stackrel{(c)}{=}\mathbb{E}_{\Phi}\Bigg[\!\prod_{i\in\Phi}\!\sum_{n=0}^{N}\!\!\Bigg(\underbrace{\int_{\mathbb{R}^{+}}\!\int_{0}^{2\pi}\!\mathcal{K}_{1}(x_1,x_2,r,\theta)f_{R}(r)f_{\theta}(\theta)d\theta dr\!}_{\mathcal{K}_{2}(x_1,x_2)}\Bigg)^n \mathbb{P}(n|n\leq N)\Bigg] \nonumber \\	&\stackrel{(d)}{=}\!\exp\!\left(-\lambda_{c}\int_{\mathbb{R}^2}\!\left(1\!-\!\!\sum_{n=0}^{N}\!\left(\mathcal{K}_{2}(x_1,x_2)\right)^n  \mathbb{P}(n|n\!\leq \!N)\right)dx\right),
	%&\stackrel{(e)}{=}e^{\!\Bigg(\!-\lambda_{c}\!\bigintsss_{\mathbb{R}^+}\bigintsss_{0}^{2\pi}\!\Big(\!1\!-\!\sum_{n=0}^{N}\mathcal{K}_{2}(l\cos\phi,l\sin \phi)^n \mathbb{P}(n|n \leq N)\!\Big) l d\phi dl\Bigg)}, 
\end{align}
where (a) follows the Gamma distribution of Nakagami fading channel gains, (b) and (d) are based on the probability generating functional (PGFL) of a PPP \cite{chiu2013stochastic}. 
In (c), we use the fact that the number of corridors is Poisson distributed conditioned on total being less than $N$.
By converting from Cartesian to polar coordinates with $x_{1} = l\cos\phi$ and $x_{2} = l\sin \phi$ in (\ref{lemma_results}), we can the final results in (\ref{(4)}).

\subsection{Proof of Theorem \ref{theorem}}
\label{appendix_B}
The connectivity probability can be calculated as follows
\begin{align}
	\mathbb{P}_\text{conn}&= \mathbb{E}\left[\mathbb{P}\left(\frac{g (h^2 + \beta^2)^{-\frac{\alpha}{2}}}{I}\geq\gamma\lvert \beta\right)\right]	
	= \mathbb{E}\left[\mathbb{P}\left(g \geq \frac{\gamma I}{(h^2 + \beta^2)^{-\frac{\alpha}{2}}}|\beta\right)\right]	 \nonumber \\ 
	&\stackrel{(a)}{\approx}1- \mathbb{E}\left[\left(1-\exp\left(\frac{-\eta\gamma I}{(h^2 + \beta^2)^{-\frac{\alpha}{2}}}\right)\right)^{m}|\beta\right]   \nonumber \\ 
	&\stackrel{(b)}{=}\sum_{\hat{m}=1}^{m}(-1)^{\hat{m}+1}{m\choose \hat{m}}\mathbb{E}\left[\exp\left(\frac{-\hat{m}\eta\gamma I}{(h^2 + \beta^2)^{-\frac{\alpha}{2}}}\right)|\beta\right] \nonumber \\ 
	&\stackrel{(c)}{=}\int_{\mathbb{R}^{+}}\sum_{\hat{m}=1}^{m}(-1)^{\hat{m}+1}{m\choose \hat{m}} \mathcal{L}(\hat{m}\eta\gamma(h^2+\beta^2)^{\frac{\alpha}{2}})f_{B}(\beta)d\beta, 
\end{align}
where (a) is based on the approximated tail probability of a Gamma function \cite{6932503}, (b) follows the Binomial theorem and the assumption that $m$ is an integer, and (c) follows the definition of Laplace transform of interference and the fact that the distance between the aircraft and GBS is a random variable.

\subsection{Proof of Lemma \ref{circular}}
\label{Appendix_C}
Consider the CP for the arbitrarily selected corridor is $(x_{1},x_{2})$ and the relative location of corridor to the CP is determined by the distance $r$ and the angle $\theta \in [0,2\pi)$ between the positive $x$-axis and the line passing through the CP and being perpendicular to the corridor.
In this case, we can derive that any point $(y_{1},y_{2})$ on the corridor will meet the following equality constraint: 
\begin{align}
	y_{2} - \tan \hat{\theta}y_{1} - (x_{2}+r\sin\theta) + \tan \hat{\theta}(x_{1}+r\cos\theta)=0, 
\end{align}
where $\hat{\theta}=	
\begin{cases}
	\theta + \frac{\pi}{2}, \quad &\text{if} \, \theta \in [0,\frac{\pi}{2}), \\
	\theta - \frac{\pi}{2}, \quad &\text{if} \, \theta \in [\frac{\pi}{2},\frac{3\pi}{2}), \\
	\theta - \frac{3\pi}{2}, \quad &\text{if} \, \theta \in [\frac{3\pi}{2},2\pi). \\
\end{cases}$ The distance between the corridor and the center of the circular area can be calculated as \cite{spain2007analytical}
\begin{align}
	d = \frac{|\tan \hat{\theta}(x_{1}+r\cos \theta)-(x_{2}+r\sin \theta)|}{\sqrt{1+(\tan^2 \hat{\theta})}} = |\cos \hat{\theta}||\tan \hat{\theta}(x_{1}+r\cos \theta)-(x_{2}+r\sin \theta)|. 
\end{align}
Hence, the length of corridor can be expressed as 
\begin{align}
	\label{r1}
	&L =2 \sqrt{R^2-d^2} = 2 \sqrt{R^2- \mathcal{T}_{1}}, 
\end{align}
where 
\begin{align}
	\label{r2}
	\mathcal{T}_{1} = &\sin^2 \hat{\theta}(x_{1}^2+ 2x_{1}r\cos \theta + r^2 \cos^2 \theta)-2\sin \hat{\theta}\cos \hat{\theta}(x_{1}x_{2}+x_{1}r\sin \theta + x_{1}r\sin \theta + x_{2}r\cos \theta \nonumber \\ &+ r^2 \sin \theta \cos \theta) + \cos^2 \hat{\theta}(x_2^2 + 2 x_{2} r \cos \theta + r^2 \cos^2 \theta) \nonumber \\ 
	\stackrel{(a)}{=}&\cos^2 \theta(x_{1}^2+ 2x_{1}r\cos \theta + r^2 \cos^2 \theta)-2\sin \theta\cos \theta(x_{1}x_{2}+x_{1}r\sin \theta + x_{1}r\sin \theta + x_{2}r\cos \theta \nonumber \\ &+ r^2 \sin \theta \cos \theta) + \sin^2 \theta(x_2^2 + 2 x_{2} r \cos \theta + r^2 \cos^2 \theta) \nonumber \\
	\stackrel{(b)}{=}&x_{1}^2 \cos^2 \theta + 2 x_1 r \cos^3 \theta + r^2 \cos^{4} \theta + 2 x_{1}x_{2}\sin \theta  \cos \theta + 2 x_{1}r \sin^2 \theta \cos \theta + 2 x_{2} r \sin \theta \cos^2\theta   \nonumber \\
	&+ 2 r^2 \sin^2 \theta \cos^2 \theta + \sin^2 \theta x_{2}^{2} + 2 x_{2}r \sin^3 \theta + r^2 \sin^4 \theta \nonumber \\ 
	=& x_{1}^2 \cos^{2}\theta + 2 x_1 r \cos \theta + r^2 + 2 x_{1}x_{2}\sin\theta \cos \theta + x^{2}_{2}\sin^{2}\theta + 2 x_{2} r \sin \theta \nonumber \\ 
	=&(x_1 \cos \theta + x_{2} \sin \theta+r)^2,   
\end{align}
where in (a), we use the relationship between $\theta$ and $\hat{\theta}$, and in (b), we simplify the results by using the basic trigonometry properties, e.g., $\cos^2 \theta + \sin^2 \theta = 1$.
After substituting (\ref{r2}) into (\ref{r1}), the length of the corridor intercepting the circular area with radius $R$ can be calculated. 
Based on the distribution of $r$ and $\theta$, we can obtain the expected length as shown in (\ref{r}).

\subsection{Proof of Theorem \ref{theorem_2}}
\label{Appendix_D}
The expected number of aircraft associated to each GBS is 
\begin{align}
	W =& \mathbb{E}\left(\frac{W_{1}\sum_{i\in\Phi}\lambda_{t} L}{W_{2}}\right) 
	=\mathbb{E}\left(\frac{W_{1}\sum_{i\in\Phi}  \lambda_{t} \mathbb{E}(L)}{W_{2}}\right) \nonumber \\ 
	=&\mathbb{E}\left(\frac{W_{1}	\sum_{n=0}^{N} \lambda_{t}  \mathbb{E}(L)n\mathbb{P}(n|n<N)  }{W_{2}}\right) \nonumber \\ 
	=& \left(\frac{\mathbb{E}(W_{1})	\sum_{n=0}^{N} \lambda_{t}  \mathbb{E}(L)n\mathbb{P}(n|n<N)  }{\mathbb{E}(W_{2})}\right) \nonumber \\ 
	\stackrel{(a)}{=}& \left(\frac{\lambda_{c} \pi R^2 	\sum_{n=0}^{N} \lambda_{t}  \mathbb{E}(L)n\mathbb{P}(n|n<N)  }{\lambda_{b} \pi R^2} 
	\right),
\end{align}
where $W_{1}$ and $W_{2}$, respectively, represent the total number of CPs and GBSs. In (a), we use the basic Poisson distribution property and the fact that the density of two-dimensional PPP distributed of CPs and GBSs will be the unit density times the area of the circular space. 

\subsection{Proof of Theorem \ref{theorem_3}}
\label{Appendix_E}
To prove Theorem \ref{theorem_3}, we first obtain the upper bound of $f(\boldsymbol{w})$: 
\begin{align}
	\label{proof_theorem3_1}
	f(\boldsymbol{w}_{i+1}) &\stackrel{(a)}{\leq} f(\boldsymbol{w}_{i}) + \langle\nabla f(\boldsymbol{w}_{i}), \boldsymbol{w}_{i+1}-\boldsymbol{w}_{i}\rangle + \frac{1}{2!}(\boldsymbol{w}_{i+1}-\boldsymbol{w}_{i})^T \nabla f(\boldsymbol{w}_{i})(\boldsymbol{w}_{i+1}-\boldsymbol{w}_{i}) \nonumber \\ 
	&\stackrel{(b)}{\leq}f(\boldsymbol{w}_{i}) + \langle\nabla f(\boldsymbol{w}_{i}), \boldsymbol{w}_{i+1}-\boldsymbol{w}_{i}\rangle + \frac{L}{2}||\boldsymbol{w}_{i+1}-\boldsymbol{w}_{i}||^2 \nonumber \\ 
	&\stackrel{(c)}{\leq}f(\boldsymbol{w}_{i}) + \langle \nabla f(\boldsymbol{w}_{i}), -\eta_{i}g_{1}(\delta_{k,i})\frac{s_{k}}{s_{K}}\nabla f_{k}(\boldsymbol{w}_{\delta_{k,i}}) \rangle + \frac{L}{2}\eta^{2}_{i}g_{1}^{2}(\delta_{k,i})(\frac{s_{k}}{s_{K}})^2 || \nabla f_{k}(\boldsymbol{w}_{\delta_{k,i}}) ||^2,
\end{align}
where (a) follows the Taylor expansion, (b) is based on the assumption of the Lipschitz continuity, and the inequality is derived due to the basic relationship between $\boldsymbol{w}_{i+1}$ and $\boldsymbol{w}_{i}$.

In the results (\ref{proof_theorem3_1}), we need to consider two types of randomness, i.e., the aircraft $k\in \mathcal{K}$ participated into the learning model parameters update and the staleness value $\delta_{k,i}$. 
First, we take expectation for both side of (\ref{proof_theorem3_1}) in terms of the participating aircraft as follows 
\begin{align}
	\label{23}
	\mathbb{E}_{k}(f(\boldsymbol{w}_{i+1})) \leq& f(\boldsymbol{w}_{i})- \eta_{i}g_{1}(\delta_{k,i})\frac{1}{K} 
	\langle \nabla f(\boldsymbol{w}_{i}), \mathbb{E}( \nabla f_{k}(\boldsymbol{w}_{\delta_{k,i}})) \rangle
	+\frac{L}{2}\eta^{2}_{i}g_{1}^{2}(\delta_{k,i})\frac{1}{K^2} || \nabla f_{k}(\boldsymbol{w}_{\delta_{k,i}}) ||^2 \nonumber \\
	\stackrel{(a)}{\leq}&f(\boldsymbol{w}_{i})- \eta_{i}g_{1}(\delta_{k,i})\frac{1}{K} 
	\langle \nabla f(\boldsymbol{w}_{i}),  \nabla f(\boldsymbol{w}_{\delta_{k,i}}) \rangle
	+\frac{L}{2}\eta^{2}_{i}g_{1}^{2}(\delta_{k,i})\frac{1}{K^2} || \nabla f_{k}(\boldsymbol{w}_{\delta_{k,i}}) ||^2 \nonumber \\
	\stackrel{(b)}{\leq}&f(\boldsymbol{w}_{i})- \eta_{i}g_{1}(\delta_{k,i})\frac{1}{2K} 
	\left( || \nabla f(\boldsymbol{w}_{i})||^2 + ||\nabla f(\boldsymbol{w}_{\delta_{k,i}}) ||^2 - ||\nabla f(\boldsymbol{w}_{i})-\nabla f(\boldsymbol{w}_{\delta_{k,i}})||^2 \right)
	\nonumber \\ &+\frac{L}{2}\eta^{2}_{i}g_{1}^{2}(\delta_{k,i})\frac{1}{K^2} || \nabla f_{k}(\boldsymbol{w}_{\delta_{k,i}}) ||^2 \nonumber \\
	=& f(\boldsymbol{w}_{i})- \frac{\eta_{i}g_{1}(\delta_{k,i})}{2K} 
	\left( || \nabla f(\boldsymbol{w}_{i})||^2 + ||\nabla f(\boldsymbol{w}_{\delta_{k,i}}) ||^2\right) \nonumber \\ &+ \frac{\eta_{i}g_{1}(\delta_{k,i})}{2K} \left(||\nabla f(\boldsymbol{w}_{i})-\nabla f(\boldsymbol{w}_{\delta_{k,i}})||^2 \right)
	+\frac{L\eta^{2}_{i}g_{1}^{2}(\delta_{k,i})}{2K^2}  ||\nabla f_{k}(\boldsymbol{w}_{\delta_{k,i}}) ||^2 \nonumber \\ 
	\stackrel{(c)}{\leq}& f(\boldsymbol{w}_{i})- \frac{\eta_{i}g_{1}(\delta_{k,i})}{2K} 
	\left( || \nabla f(\boldsymbol{w}_{i})||^2 + ||\nabla f(\boldsymbol{w}_{\delta_{k,i}}) ||^2\right) \nonumber \\
	&+ \frac{L\eta^{2}_{i}g_{1}^{2}(\delta_{k,i})}{2K^2} g_{2}(\delta_{k,i}) \mathbb{E}[||\nabla f(\boldsymbol{w}_{\delta_{k,i}})||^2] +  \frac{L\eta^{2}_{i}g_{1}^{2}(\delta_{k,i})}{2K^2}  \underbrace{||\nabla f_{k}(\boldsymbol{w}_{\delta_{k,i}}) ||^2}_{\mathcal{T}_{2}},
\end{align}
where in (a) follows the fact that, for aircraft $k \in \mathcal{K}$, the gradient of its local training data is the unbiased estimation to the gradient representation for the data across all aircraft, i.e., $\mathbb{E}(f_{k}(\boldsymbol{w})) = \mathbb{E}(f(\boldsymbol{w}))$, $\forall k \in \mathcal{K}$ \cite{chen2019fast}. The changes in (b) are based on $\langle \boldsymbol{w}, \boldsymbol{\hat{w}_{1}}\rangle = \frac{1}{2}(||\boldsymbol{w}||^2 + ||\boldsymbol{\hat{w}_{1}}||^2 - ||\boldsymbol{w}-\boldsymbol{\hat{w}_{1}}||^2 )$. 
In (c), we use the third assumption regarding the variance between the global gradient descent at an arbitrarily communication round $i$ and the outdated local gradient at a random aircraft.
In particular, $\mathcal{T}_{2}$ can be further simplified as 
\begin{align}
	\label{24}
	\mathcal{T}_{2} &= \mathbb{E}_{k}(||\nabla f_{k}(\boldsymbol{w}_{\delta_{k,i}})- f(\boldsymbol{w}_{\delta_{k,i}}) + f(\boldsymbol{w}_{\delta_{k,i}})||^2) \nonumber \\ 
	&=  \mathbb{E}_{k}(||\nabla f_{k}(\boldsymbol{w}_{\delta_{k,i}})- f(\boldsymbol{w}_{\delta_{k,i}})||^2) +  \mathbb{E}_{k}(||f(\boldsymbol{w}_{\delta_{k,i}})||^2) + 2 \mathbb{E}_{k} \langle\nabla f_{k}(\boldsymbol{w}_{\delta_{k,i}})- f(\boldsymbol{w}_{\delta_{k,i}}), f(\boldsymbol{w}_{\delta_{k,i}}) \rangle 
	\nonumber \\
	&= \mathbb{E}_{k}(||\nabla f_{k}(\boldsymbol{w}_{\delta_{k,i}})- f(\boldsymbol{w}_{\delta_{k,i}})||^2) +  \mathbb{E}_{k}(||f(\boldsymbol{w}_{\delta_{k,i}})||^2) \nonumber \\ 
	&\stackrel{(a)}{\leq} \phi^2  + \mathbb{E}(||f(\boldsymbol{w}_{\delta_{k,i}})||^2),
\end{align}
where in (a), we use the second assumption. After replacing $\mathcal{T}_{2}$ with the results obtained in (\ref{24}), we can further simplify (\ref{23}) as 
\begin{align}
	\label{25}
	\mathbb{E}_{k}(f(\boldsymbol{w}_{i+1})) \leq& f(\boldsymbol{w}_{i})- \frac{\eta_{i}g_{1}(\delta_{k,i})}{2K} 
	\left( || \nabla f(\boldsymbol{w}_{i})||^2 + ||\nabla f(\boldsymbol{w}_{\delta_{k,i}}) ||^2\right) \nonumber \\
	&+ \frac{L\eta^{2}_{i}g_{1}(\delta_{k,i})}{2K^2} g_{2}(\delta_{k,i}) \mathbb{E}[||\nabla f(\boldsymbol{w}_{\delta_{k,i}})||^2] +  \frac{L\eta^{2}_{i}g_{1}^{2}(\delta_{k,i})}{2K^2}  \left(\phi^2  + \mathbb{E}(||f(\boldsymbol{w}_{\delta_{k,i}})||^2)\right) \nonumber \\ 
	=&f(\boldsymbol{w}_{i})- \frac{\eta_{i}g_{1}(\delta_{k,i})}{2K} \left( ||\nabla f(\boldsymbol{w}_{i})||^2\right)+\frac{L\eta^{2}_{i}g_{1}^{2}(\delta_{k,i})\phi^2}{2K^2} \nonumber \\ &+ \left(\frac{L\eta^{2}_{i}g_{1}(\delta_{k,i})}{2K^2} g_{2}(\delta_{k,i})+\frac{L\eta^{2}_{i}g_{1}^{2}(\delta_{k,i})}{2K^2}  -  \frac{\eta_{i}g_{1}(\delta_{k,i})}{2K} \right)  \mathbb{E}[||\nabla f(\boldsymbol{w}_{\delta_{k,i}})||^2]. 
\end{align}
Taking expectation of (\ref{25}) in terms of the staleness, we have 
\begin{align}
	\label{(26)}
	\mathbb{E}_{k,\delta}(f(\boldsymbol{w}_{i+1})) \leq& f(\boldsymbol{w}_{i})- \frac{\eta_{i}\mathbb{E}_{\delta}(g_{1}(\delta_{k,i}))}{2K} \left( ||\nabla f(\boldsymbol{w}_{i})||^2\right)+\frac{L\eta^{2}_{i}\mathbb{E}_{\delta}(g_{1}^{2}(\delta_{k,i}))\phi^2}{2K^2}\nonumber \\ &+ \mathbb{E}_{\delta}\left(\left(\frac{L\eta^{2}_{i}g_{1}(\delta_{k,i})}{2K^2} g_{2}(\delta_{k,i})+\frac{L\eta^{2}_{i}g_{1}^{2}(\delta_{k,i})}{2K^2}  -  \frac{\eta_{i}g_{1}(\delta_{k,i})}{2K} \right)  \mathbb{E}[||\nabla f(\boldsymbol{w}_{\delta_{k,i}})||^2]\right). \nonumber 
\end{align}
If $L\eta_{i}g_{2}(\delta_{k,i}) +L\eta_{i}g_{1}(\delta_{k,i})  -  K \leq 0, \forall \delta_{k,i} \in \mathbb{R}_{+}$, we can obtain the convergence rate as presented in Theorem \ref{theorem_3}.

\subsection{Proof of Corollary \ref{corollary2}}
\label{Appendix_F}
After subtracting $f(\boldsymbol{w}^{*})$ in both sides of (\ref{13}), we have 
\begin{align}
	&\mathbb{E}(f(\boldsymbol{w}_{i+1}))\!-\!f(\boldsymbol{w}^{*})\! \leq f(\boldsymbol{w}_{i})\! - \!f(\boldsymbol{w}^{*})\! -\! \frac{\eta_{i}\mathbb{E}_{\delta}(g_{1}(\delta))}{2K}  ||\nabla f(\boldsymbol{w}_{i})||^2\!+\!\frac{L\eta^{2}_{i}\left(\mathbb{V}_{\delta}(g_{1}(\delta))\!+\!\mathbb{E}_{\delta}^2(g_{1}(\delta))\right)\phi^2}{2K^2}\nonumber \\ 
	&\stackrel{(a)}{\leq}f(\boldsymbol{w}_{i}) - f(\boldsymbol{w}^{*}) - \frac{\eta_{i}\mathbb{E}_{\delta}(g_{1}(\delta))c}{K} \left( f(\boldsymbol{w}_{i}) - f(\boldsymbol{w}^{*})\right) +\frac{L\eta^{2}_{i}\left(\mathbb{V}_{\delta}(g_{1}(\delta))+\mathbb{E}_{\delta}^2(g_{1}(\delta))\right)\phi^2}{2K^2}\nonumber \\ 
	&=\left(1-\frac{\eta_{i}\mathbb{E}_{\delta}(g_{1}(\delta))c}{K}\right)\left( f(\boldsymbol{w}_{i}) - f(\boldsymbol{w}^{*})\right)+\frac{L\eta^{2}_{i}\left(\mathbb{V}_{\delta}(g_{1}(\delta))+\mathbb{E}_{\delta}^2(g_{1}(\delta))\right)\phi^2}{2K^2} \nonumber \\ 
	&=\left(1-\frac{\eta_{i}\mathbb{E}_{\delta}(g_{1}(\delta))c}{K}\right)^2\left( f(\boldsymbol{w}_{i-1}) - f(\boldsymbol{w}^{*})\right) +
	\frac{L\eta^{2}_{i}\left(\mathbb{V}_{\delta}(g_{1}(\delta))+\mathbb{E}_{\delta}^2(g_{1}(\delta))\right)\phi^2}{2K^2} \nonumber \\ &+ \left(1-\frac{\eta_{i}\mathbb{E}_{\delta}(g_{1}(\delta))c}{K}\right) \left(\frac{L\eta^{2}_{i}\left(\mathbb{V}_{\delta}(g_{1}(\delta))+\mathbb{E}_{\delta}^2(g_{1}(\delta))\right)\phi^2}{2K^2}\right)\nonumber \\ 
	& ... \nonumber \\
	&= \left(1-\frac{\eta_{i}\mathbb{E}_{\delta}(g_{1}(\delta))c}{K}\right)^{i+1}\left( f(\boldsymbol{w}_{0}) - f(\boldsymbol{w}^{*})\right) \nonumber \\ &+ \sum_{j=0}^{i}\left(1-\frac{\eta_{i}\mathbb{E}_{\delta}(g_{1}(\delta))c}{K}\right)^j \frac{L\eta^{2}_{i}\left(\mathbb{V}_{\delta}(g_{1}(\delta))+\mathbb{E}_{\delta}^2(g_{1}(\delta))\right)\phi^2}{2K^2},
\end{align}
where in (a), we use the lower bound for the norm $f(\boldsymbol{w})$, i.e., $||\nabla f(\boldsymbol{w}_{i})||^2\geq 2c (f(\boldsymbol{w}_{i}) - f(\boldsymbol{w}^{*}))$ \cite{zeng2020ICC}.
Based on the basic property of geometric progression, we can have the simplified results in Corollary \ref{corollary2}.
\def\baselinestretch{1.05}
\bibliographystyle{IEEEtran}

\end{document}